\documentclass[english]{article}
\usepackage[T1]{fontenc}
\usepackage[latin9]{inputenc}
\usepackage{babel}
\usepackage{graphicx,amsmath,amssymb,color}
\usepackage[normalem]{ulem}
\usepackage{amsfonts}
\usepackage[toc,page]{appendix}
\usepackage{hyperref}
\usepackage{latexsym}
\usepackage{amsfonts}
\usepackage{algpseudocode}
\usepackage{amsthm}
\usepackage{mathrsfs}
\usepackage{color,verbatim}
\usepackage{psfrag}
\usepackage{rotating} 

\usepackage{subcaption} 

\usepackage[svgnames]{xcolor}
\usepackage{tikz}

\usepackage{algorithm}

\newcommand{\bra}[1]{\langle#1 |}
\newcommand{\ket}[1]{|#1 \rangle}

\newcommand{\bigbraket}[2]{\left \langle #1 \middle\vert #2 \right \rangle}

\newcommand{\sandwich}[3]{\left \langle #1 \middle \vert #2 \middle \vert #3 \right\rangle}

\definecolor{myDarkRed}{rgb}{0.5, 0, 0}

\definecolor{myOrange}{rgb}{0.7, 0.3, 0} 

\definecolor{myDarkGreen}{rgb}{0, 0.7, 0} 

\definecolor{myAqua}{rgb}{0, 0.3, 0.3}

\begin{document}

\title{Suppressing unwanted fluctuations in QAOA and approximate quantum annealing}
\author{Touheed Anwar Atif, Catherine Potts, David Haycraft,\\ Raouf Dridi, and Nicholas Chancellor}
\date{Quantum Computing Inc.~Leesburg, VA, USA \\ \today}

\maketitle

\begin{abstract}

The quantum approximate optimisation algorithm (QAOA) was partially inspired by digitising quantum annealing. Based on this inspiration, we develop techniques to use the additional flexibility of a universal gate-model quantum computer to mitigate fluctuation effects which are known to distort the search space within quantum annealing and lead to false minima. We find that even just the added ability to take Pauli X measurements allows us to modify the mixer angles to counteract these effects by scaling mixer terms in a way proportional to the diagonal elements of the Fubini-Study metric. We find that mitigating these effects can lead to higher success probabilities in cases where the energy landscape is distorted and that we can use the same Pauli X measurements to target which variables are likely to be susceptible to strong fluctuations. The effects of the methods we introduce are relevant even at relatively low depth of $p=10-20$, suggesting that the techniques we are developing are likely to be relevant in the near term. Furthermore, since these methods rely on controlling a degree of freedom which is not typically modified in QAOA, our methods will be compatible with a wide range of other QAOA innovations. We further verify that these fluctuation effects can be observed on an IonQ Harmony QPU.
\end{abstract}

\section{Introduction}

Solving combinatorial optimisation problems on near term gate-model quantum computers poses an interesting challenge. It was shown by Grover \cite{Grover1996search} that quantum computers have a provable advantage over any possible quantum technique in searching an unstructured search space. Furthermore it is possible to leverage this advantage as a subroutine in algorithms for solving real optimisation problems (which do have structure to exploit) for example \cite{Monatanaro2020branchandbound}. While having the advantage of being more amenable to proofs of scaling, algorithms of this type have the disadvantage that a call to a Grover subroutine would have a depth which scales as $\sqrt{N}$, where $N$ is the size of the space which the subroutine searches, and will typically be exponential in the size of the problem (though possibly much smaller than the total size of the solution space if used in a subroutine). Since near term devices are only going to be able to implement algorithms of limited depth due to noise, it is important to consider heuristic techniques which do not require deep circuits to operate.

Variational techniques \cite{Cerezo2021variational,Callison2022hybrid} which involve the use of classical feedback to optimise a circuit to output high quality solution candidates are one family of techniques which are commonly used in this direction. The two most popular variational techniques are the variational quantum eigensolver (VQE) \cite{Peruzzo2014VQE}, and the quantum approximate optimisation algorithm\cite{Farhi14a} (QAOA, alternate acronym quantum alternating operator ansatz \cite{Hadfield2019QAOA}). In this paper we focus on the latter, which has a structure similar to a digitized simulation of the analog quantum annealing metaheuristic. Given this similar structure, we also consider an additional algorithm, known as approximate quantum annealing (AQA) \cite{Willsch2022AQA}, which is a non-variational algorithm based on choosing parameters which simulate a quantum anneal. While Willsch et.~al.\cite{Willsch2022AQA} coined the term AQA, a variety of previous works have considered similar protocols \cite{Zhou2020AQA,Streif2020AQA,Sack2021quantumannealing,streif2019comparison}, while not explored here, \cite{Sack2021quantumannealing} makes the important observation that AQA can be used as an initialisation for QAOA optimisation, this is further tested in \cite{Willsch2022AQA}. A related but distinct approach is described in \cite{Kolotouros22PQC}.

A key advantage of QAOA and AQA over analog quantum annealing is that measurements can be performed in different bases and quantities which would not typically be accessible can be used to modify the implementation. In this work we explore one way in which these measurements can be used, which has a  connection to the underlying geometry of the manifold of quantum states i.e., the complex projective space of rays, through the Fubini-Study metric \cite{Fubini1904metric,Study1905metric}. In particular we relate the value of diagonal elements of this metric to the amount of useful mixing within the state space which single gates within the mixer are able to do.

Our use of the Fubini-Study metric differs from previous uses for example quantum natural gradients, where the pseudo-inverse of the Fubini-Study metric is employed as geometric preconditioning of the gradient descent update \cite{Stokes2020quantumnatural}.

Recent investigations \cite{Wierichs_2020, touheed}  suggest this geometry-aware preconditioning outperforms other, gradient and non gradient based, optimisation methods (with the overhead of computing second order derivatives).

By building on problems statements which are known to behave pathologically from  quantum annealing literature \cite{dickson13a,Boixo2013a} we are able to demonstrate the efficacy of these methods.

\section{Background}

\subsection{QAOA and AQA}

QAOA is a variational quantum heuristic which is implemented by applying a \textbf{phase separator} and a \textbf{mixer} operation in a sequential manner\cite{Farhi14a}. The level of rotation applied in each operation is then optimised variationally. A related, non-variational heuristic is known as AQA\cite{Willsch2022AQA} and involves applying these terms sequentially in a way which simulates a continuous time quantum anneal, possibly optimising over the total rotation. Since the structure of the circuits to implement these algorithms is the same, we will first discuss the structure of the mixer and phase separator components and then return to discussion of these algorithms.

The phase separator defines a problem which the user desires to solve (usually a hard combinatorial optimisation problem). The problem is encoded such that phase rotation is applied in proportion to how optimal computational basis states are but does not cause transitions between computational basis states. In practice, this is usually achieved using one- and two-qubit gates, which are diagonal in the $Z$ basis, $Z$ and controlled-$Z$ rotation gates. Since we use the standard construction for the phase separator, we will not review it here, however since we use a modified mixer we will discuss that construction in detail, we call the phase separator $U_{\mathrm{phase}}(Q,\gamma)$.  
In the simplest case, a mixer unitary can be defined using powers of Pauli X operations as 
\begin{equation}
U_{\mathrm{mix}}(\beta)=\prod_{j}X_j^{-\frac{\beta}{\pi}}.
\end{equation}
For the purposes of this study, we consider a slightly more general mixer, which allows the rotations to be applied differently to different qubits:

\begin{equation}
U_{\mathrm{mix}}(\beta,\vec{\zeta})=\prod_{j}X_j^{-\frac{\beta}{\pi}\zeta_j}. \label{eq:mixer_zeta}
\end{equation} To make calculations easier later in the paper, we note that because $X_j=-i\exp(i \pi X_j)$, this unitary can also be written as: 

\begin{equation}
U_{\mathrm{mix}}(\beta,\vec{\zeta})=(-i)^n\prod_{j}\exp(-i\beta\zeta_jX_j)=(-i)^n\exp\left(-i\beta\sum_j\zeta_jX_j\right),\label{eq:mixer_exp}
\end{equation} from which we can drop the irrelevant global phase which precedes the product. A similar exponential representation exists for the phase separator but we do not derive it here because we do not use it in further calculations. 

A single stage of the QAOA (or AQA) protocol than consists of application of the phase separator followed by the mixer:
\begin{equation}
    U(\gamma,\beta,\vec{\zeta})=U_{\mathrm{phase}}(Q,\gamma)U_{\mathrm{mix}}(\beta,\vec{\zeta}).
\end{equation}
In traditionally formulated QAOA (AQA), the mixer angles are all the same across qubits, and this simplifies to: 
\begin{equation}
    U(\gamma,\beta)=U_{\mathrm{phase}}(Q,\gamma)U_{\mathrm{mix}}(\beta,\vec{1}),\label{eq:traditional_layer}
\end{equation}
where we use $\vec{1}$ as shorthand for a vector of the appropriate length consisting of all $1$ values.

The total protocol then becomes: 
\begin{equation}
    U_{\mathrm{Seq}}(\vec{\gamma},\vec{\beta},\vec{\zeta})=\prod_{l=0}^{p-1}U(\gamma_l,\beta_l,\vec{\zeta}^l),
\end{equation}
where $p$ is the total number of stages of the protocol. QAOA and AQA apply the same form of unitary, but differ in how $\vec{\gamma}$  and $\vec{\beta}$ are chosen.\footnote{Unlike \cite{Willsch2022AQA}, which uses $n$ for the number of applications of mixers and phase separators in AQA and $p$ for QAOA. We have elected to use $p$ for both so that the unitary can be described by the same formula.} In QAOA they are found by iteratively applying a classical optimisation protocol, while in AQA they are chosen to match a course grained simulation of an anneal. We will discuss how $\vec{\zeta}$ could be found in the next section. Unlike AQA, QAOA requires many repetitions of the protocols to find the optimal $\vec{\gamma}$  and $\vec{\beta}$ and incurs substantial numerical overhead. 

We consider simple AQA protocol, which approximates linear anneals such that $\gamma$ increases in a linear fashion from $\frac{\tau}{p}$ to $\tau$, and $\beta$ likewise decreases. Note that to ensure that our AQA is a closer analog to a QAOA protocol, we use a first order Trotterisation, although we would expect it to find similar results for the second-order method used in \cite{Willsch2022AQA}. Expressed mathematically, for $p$ interactions we have:
\begin{align}
\gamma_l=r_l \tau, \label{eq:gamma_l}\\
\beta_l=\left(1-r_l \right) \tau, \label{eq:beta_l} \\
r_l=\frac{l+1}{p+1} \label{eq:r_l}.
\end{align}
Note that the values have been chosen such that $\gamma_l+\beta_l=\tau$, but also such that for $l\le 0 \le p-1$, we never have a ``trivial'' iteration where $\gamma_l=0$ or $\beta_l=0$.  The variable $\tau$ is a maximum angle for a single rotation; we focus on cases where $\tau$ mildly decreases with $p$. The mild $p$ dependence is useful because for scaling $O(1)<\frac{1}{\tau(p)}<O(p)$, the rotation per layer will decrease, but the total rotation will still increase with increasing $p$. The former condition guarantees that as $p\rightarrow \infty$, the AQA protocol will become an increasingly faithful simulation of a quantum anneal, while the latter condition guarantees that the time for which the simulated system is annealed will tend toward infinity along with $p$. By the adiabatic theorem of quantum mechanics, it follows that the probability of finding an optimal solution will tend toward $1$ as $p$ tends toward $\infty$, a theoretical guarantee which does not exist for fixed $\tau$. While there are many functional dependencies which could be chosen, we choose a relatively simple polynomial dependence: $\tau(p)=\frac{\pi}{2p^{0.25}}$. 
Now that we have defined our AQA protocol, it is worth discussing the new innovations we add here. First, we review the Fubini-Study metric in the context of our work, and then we discuss how it is used to construct a protocol to modify the drive.

\subsection{Fubini-Study metric}

Following the definition given in \cite{Stokes2020quantumnatural}, we define the quantum geometric tensor as: 
\begin{equation}
G_{j,k}(\vec{\zeta})=\bigbraket{\frac{\partial \psi(\vec{\zeta})}{\partial \zeta_j}}{\frac{\partial \psi(\vec{\zeta})}{\partial \zeta_k}}-\bigbraket{\frac{\partial \psi(\vec{\zeta)}}{\partial \zeta_j}}{\psi(\vec{\zeta})}\bigbraket{\psi(\vec{\zeta})}{\frac{\partial \psi(\vec{\zeta})}{\partial \zeta_k}}, \label{eq:QGT}
\end{equation}
where $\zeta$ are different control parameters representing modification of mixer angles for individual qubits. The Fubini-Study metric is simply the real part of this quantity $F_{j,k}(\vec{\zeta})=\mathrm{Re}[G_{j,k}(\vec{\zeta})]$. Let us consider the simple case of a state operated upon by a non-uniform mixer term (defined in eq.~\ref{eq:mixer_exp} but with the global phase ignored and global factor of $\beta$ dropped):
\begin{equation}
\ket{\psi(\vec{\zeta})}=\prod_j\exp\left[-i \zeta_j X_j\right]\ket{\psi} \label{eq:X_exp},
\end{equation}
It immediately follows that: 
\begin{equation}
\frac{\ket{\partial \psi(\vec{\zeta})}}{\partial \zeta_j}=i X_j \ket{\psi(\vec{\zeta})}
\end{equation}
Plugging into equation \ref{eq:QGT} gives: 
\begin{equation}
G_{j,k}(\vec{\zeta})=\sandwich{\psi(\vec{\zeta})}{X_jX_k}{\psi(\vec{\zeta})}-\sandwich{\psi(\vec{\zeta})}{X_j}{\psi(\vec{\zeta})}\sandwich{\psi(\vec{\zeta})}{X_k}{\psi(\vec{\zeta})} \label{eq:QGT_elements},
\end{equation}
which is purely real by definition, so in this case $F_{j,k}=G_{j,k}$. These quantities can all be readily obtained through simple $X$ basis measurements of the state $\ket{\psi}$, which is equivalent to measuring $\ket{\psi(\vec{\zeta})}$ due to the fact that $\exp\left[i\beta \zeta_j X_j\right]$ will not affect $X$ basis measurements. The diagonal elements take a particularly simple form:
\begin{equation}
F_{j,j}(\vec{\zeta})=1-(\sandwich{\psi}{X_j}{\psi})^2 \label{eq:QGT_diagonal}.
\end{equation}

The physical interpretation of the Fubini-Study metric is that it acts as a Riemannian metric on the the complex projective space of pure quantum states. The expression in Eq.~\ref{eq:X_exp} defines a ray in this space (in the special case where $\ket{\psi}$ is an eigenstate, it defines the zero vector). Based on this geometric interpretation, we can understand the physical meaning of the elements of the metric. The diagonal metric elements tell us the product of this vector with itself physically; the square root of this quantity tells us how much distance is traveled if $\zeta \rightarrow \zeta+\delta \zeta$. The off-diagonal elements contain information about the distance traveled in the directions defined by the other transformations.\footnote{Note: this quantity would not simply be the off-diagonal element, since that will define the product of two un-normalized vectors in the space, to obtain this value, we would have to normalize by dividing by the square root of a diagonal element to get a product with a unit vector.}  An interesting special case is where $\ket{\psi}$ is an eigenstate of $X_j$. In this case, the action of the operation is only a global phase which is projected out in the definition of the metric. Since no transformation is induced by taking $\zeta_j \rightarrow \zeta_j+\delta \zeta$, then the diagonal and all possible off-diagonal elements of the metric corresponding to this $\zeta_j$ will be zero, which can be verified by applying Eq.~\ref{eq:QGT_elements}. 

\begin{figure}
\begin{subfigure}[b]{0.475\textwidth}
	\begin{centering}
	\includegraphics[width= \textwidth]{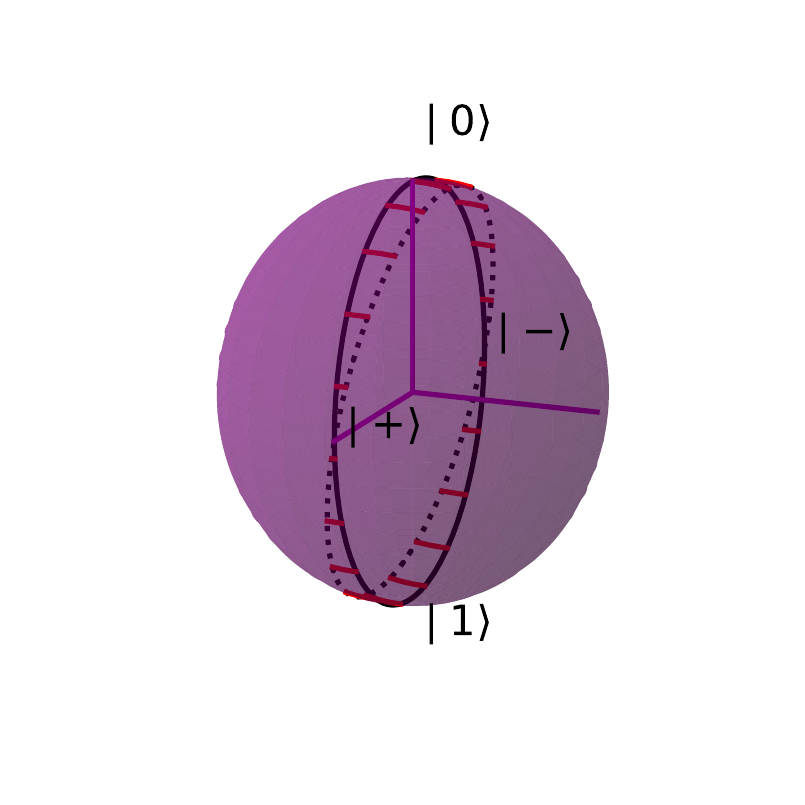}
	\par
	\end{centering}
    \caption{metric element}
    \label{fig:bloch_F}
 \end{subfigure}
 \begin{subfigure}[b]{0.475\textwidth}
	\begin{centering}
	\includegraphics[width= \textwidth]{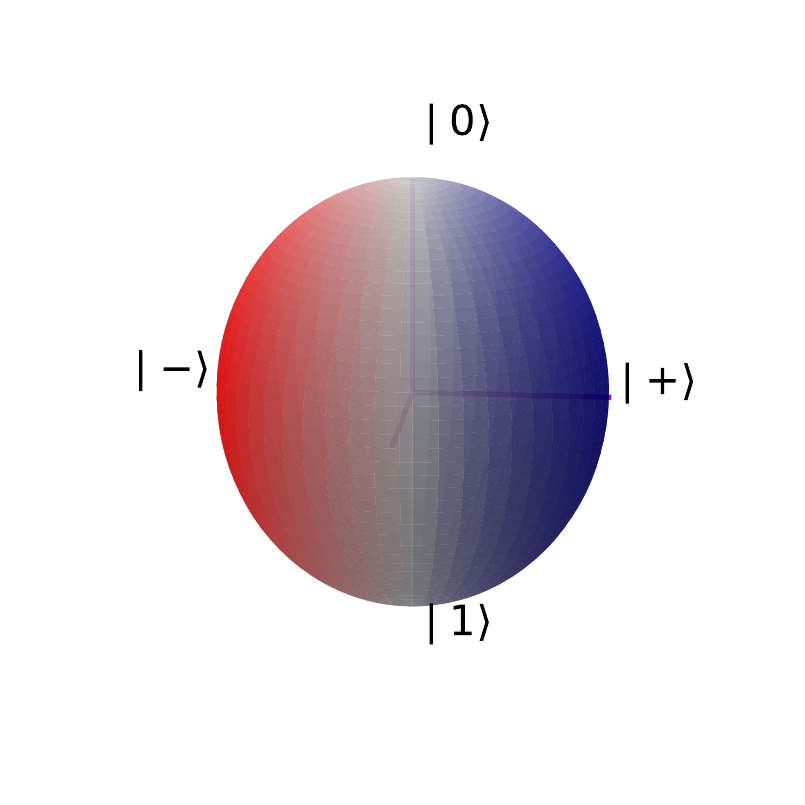}
	\par
	\end{centering}
    \caption{global phase}
    \label{fig:bloch_phase}
 \end{subfigure}
	\caption{Bloch sphere visualisation of the effect of a rotation about the $X$ axis. Left figure shows the metric element contribution: the dashed circle indicates states after rotations, while the solid circle indicates the state before (the total rotation is proportional to the square root of the diagonal metric element for small rotation angles). In particular, the $\ket{{+}}$ and $\ket{{-}}$ states experience no rotation. The right subfigure shows the global phase shift associated with this operation (note rotated for easier visualisation). More saturated blue indicates a stronger negative phase, while red indicates positive.}
	
\end{figure}

Since these represent single qubit operations, they can be visualised on the Bloch sphere. As figure \ref{fig:bloch_F} illustrates, a small rotation around the $X$ axis induces the largest rotation, and therefore corresponds to the largest diagonal metric element for states which have zero $X$ expectation.

In the special case where the starting state is an eigenstate, it is clear both mathematically and visually from figure \ref{fig:bloch_F} that no mixing between computational basis states is performed by setting $\zeta_j\neq 0$. Beyond just not being useful, this rotation can actually be harmful to  the computation. To see how this operation can be harmful, consider the case where we apply this operation to a state of the following form:
\begin{equation}
\ket{\psi}=\sqrt{1-\epsilon^2 } \ket{\psi_0}\otimes\ket{+}\otimes\ket{\psi_1}+\epsilon \ket{\psi'_0}\otimes\ket{0}\otimes\ket{\psi'_1}, \label{eq:example_psi}
\end{equation}
where: 
\begin{equation}
 \left(\bra{\psi_0}\otimes \bra{\psi_1}\right) \left(\ket{\psi'_0}\otimes\ket{\psi'_1}\right)=0.
\end{equation}
By applying Eq.~\ref{eq:QGT_diagonal}, we find that $F_{j,j}=\epsilon^2$, so the amount of mixing which this operation performs is minimal. To understand the physical effect of this operation, we examine the effect of applying Eq.~\ref{eq:X_exp} to the wave function defined in: Eq.~\ref{eq:example_psi}
\begin{align}
\exp\left[-i\beta \zeta_j X_j\right]\ket{\psi}= \nonumber \\
\sqrt{1-\epsilon^2 } \exp\left[-i\beta \zeta_j\right]\ket{\psi_0}\otimes\ket{+}\otimes \ket{\psi_1}+\nonumber \\ 
\epsilon \ket{\psi'_0}\otimes\left(\cos\left[\beta \zeta_j \right]\ket{0}-i \sin\left[\beta \zeta_j \right]\ket{1}  \right)\otimes\ket{\psi'_1}, \label{eq:example_psi}
\end{align}
Applying this driver term has two effects: the desirable effect of performing a bit flip on the component where the wave function is not an eigenstate, and an undesirable effect of applying a relative phase to the two components (see figure \ref{fig:bloch_phase}). Since this phase mimics the action of the phase separator, it will lead to phase interference which could favor a suboptimal computational basis state over the truly optimal one and therefore could globally corrupt the computation. While setting $\zeta_j=0$ in this case would completely avoid any mixing of the values of the $j$th bit, and therefore inhibit computation, one could attempt to minimize the corrupting effects by reducing the value of $\zeta_j$. In the next section, we propose a protocol for doing exactly this, based on a diagnosis from measuring the diagonal elements of the Fubini-Study metric.

\section{Protocols for modifying mixer angles}

The simple protocol we propose is to scale each $\zeta_j$ proportional to $F_{j,j}(\vec{\zeta})$. One concern is that, early in the protocol, when all qubits are close to being eigenstates of $X$, this protocol may greatly reduce the action of the mixer. Since we are actually interested in relative changes of the mixer between different qubits and not substantially modifying the overall rotation, we constrain that the maximum value of elements in $\vec{\zeta}$ to be $1$, by setting: 
\begin{equation}
\zeta_j=\frac{F_{j,j}(\vec{\zeta})}{\max_kF_{k,k}(\vec{\zeta})}.
\end{equation}
For our protocol, we estimate the $\zeta_j$ values for a given stage by using the values of $F_{j,j}$ from the previous stage. In other words, if using an index $0<l<p-1$ to index the stages of the QAOA protocol then:  
\begin{equation}
\zeta^l_j=\frac{F^{l-1}_{j,j}(\vec{\zeta})}{\max_kF^{l-1}_{k,k}(\vec{\zeta})}, \label{eq:supressed_def}
\end{equation}
for $l>0$ and $\zeta^0_j=1 \quad \forall j$. 

We additionally consider a version of the protocol where we first run the circuit without any modifications and only modify $\zeta_j$ if $F^{p-1}_{j,j}$ is less than a certain threshold for a protocol where all $\vec{\zeta}$ values are set to $1$. This modified version suppresses fluctuations only on qubits which would end the circuit in an approximate eigenstate of the $X$ Pauli operator, stated mathematically.
\begin{equation}
\zeta^l_j=\begin{cases}\frac{F^{l-1}_{j,j}(\vec{\zeta})}{\max_kF^{l-1}_{k,k}(\vec{\zeta})} & F^{p-1}_{j,j}(\vec{1})<\Theta \\ 1 & \mathrm{otherwise} \end{cases} \label{eq:thresholded_def},
\end{equation}
Recall that we use $\vec{1}$ as shorthand for a vector of ones of the appropriate length. Empirically, we find that $\Theta=0.2$ works well as a threshold, so we use that throughout the analysis presented here. For ease of communication, we refer to the protocol in Eq.~\ref{eq:supressed_def} as using a \textbf{suppressed} mixer since it is designed to suppress unwanted quantum fluctuations, and the one in Eq.~\ref{eq:thresholded_def} as using a \textbf{thresholded} mixer since suppression of fluctuations is only attempted beyond a certain threshold. We refer to the traditional QAOA protocol (and its AQA analog) where the relative strength of the individual mixer terms is not changed using an \textbf{unmodified} mixer.

\section{Problem QUBOs}

An established trick in quantum annealing experiments is to use Ising Hamiltonians, which contain ``free variables.'' These Hamiltonians are engineered to have a low energy state where many of the variables yield equal energy for the $\ket{0}$ or $\ket{1}$ configuration. In the presence of the transverse field (the annealing equivalent of the mixer we use here), the qubits corresponding to these variables are free to locally adopt the ground state of the transverse field operator. On the other hand, the true minimum is a state where all variables are strongly constrained to either the $\ket{0}$ or $\ket{1}$ configuration. This design creates a very close avoided crossing, as the system must tunnel from the low energy state to the true ground state late in the anneal. An example of how to engineer such a Hamiltonian and of the close avoided crossing created can be found in figures 3 and 4 respectively of \cite{Chancellor2021SearchRange}, or figure 2 of \cite{dickson13a}.

\begin{figure}
\begin{subfigure}[b]{0.475\textwidth}
	\begin{centering}
	\includegraphics[width=0.45 \textwidth]{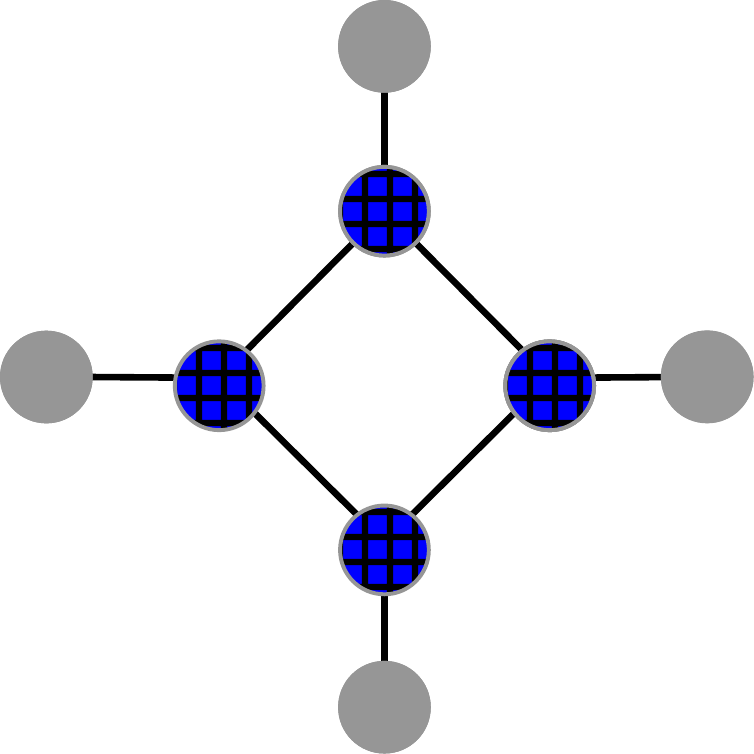}
	\par
	\end{centering}
	\caption{8 qubit}
	\label{fig:8_qubit}
\end{subfigure}
\begin{subfigure}[b]{0.475\textwidth}
	\begin{centering}
	\includegraphics[width=\textwidth]{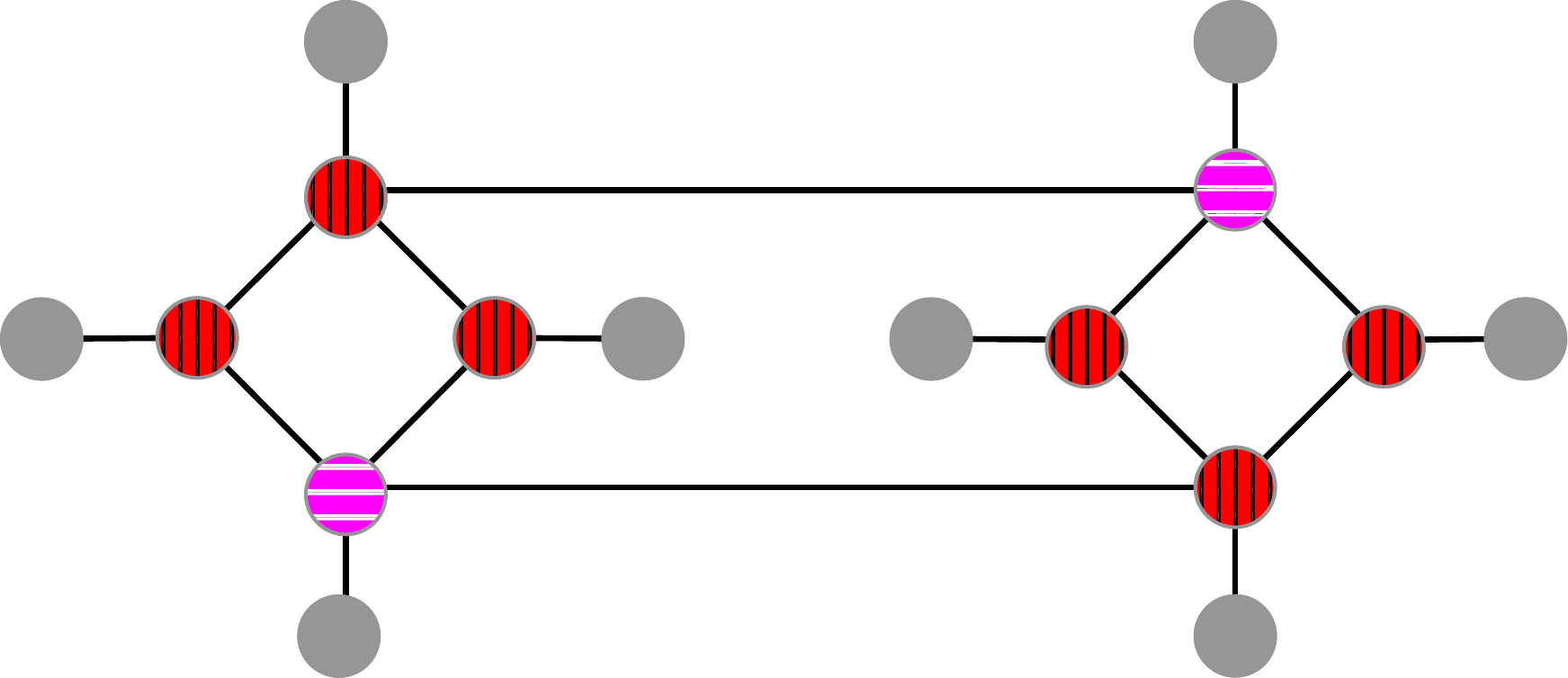}
	\par
	\end{centering}
	\caption{16 qubit}
	\label{fig:16_qubit}
\end{subfigure}
\caption{Graphical representation of 8 and $16$ qubit QUBOs engineered to create a small gap. Black lines correspond to $-1$ off diagonal elements, vertically striped red nodes correspond to qubits with a diagonal element of $+5$, while magenta horizontally striped nodes correspond to qubits with a diagonal element of $+3$ for the problem or $+4.4$ for the version with a reduced energy difference between the true and false minima.  Grey nodes correspond to qubits with a zero diagonal elements. Blue hashed filling corresponds to a diagonal element which takes a value of $3.95$.} 
\end{figure}

For our investigations, we use the 16 qubit problem from \cite{dickson13a}, which is depicted in figure \ref{fig:16_qubit}. We discover that, in contrast to quantum annealing with moderate runtimes, which results in a probability approaching $1$ to be in the false minimum, low to moderate depth AQA only results in a probability on the order of $50\%$ (see results section). We find, however, that by changing the diagonal QUBO elements in a way which reduces the difference between the false minima and the true minima by a factor of   $5$, we can create a problem where the tendency to get trapped in a false minima is much stronger. We also investigate an 8 qubit QUBO inspired by the Hamiltonian studied in \cite{Boixo2013a}, which appears in figure \ref{fig:8_qubit}.

While these problem statements contain a desirable feature for our experiments, the presence of true and false energy minima, with strong quantum fluctuations directing the anneal toward the latter, do not generally have any of the typical structure of hard optimisation problems. Particularly since they only have negative off-diagonal elements, they are not likely to have the complex landscape of false minima which is present in real, hard combinatorial optimisation problems. Fortunately, we are able to combine these Hamiltonians with features which do correspond to hard problems. To do this, we need to couple a gadget similar to the ones shown in figure \ref{fig:16_qubit} to a hard problem which is symmetric under full bit inversion. One of the two equivalent solutions to this problem will correspond to the true minimum and the other to a false minimum. A simple problem which has these desired properties is weighted maximum cut, a generalisation of the typical maximum cut problem \cite{Karp1972Reduce,Gutin2021LowerBF}, but where each edge is given a positive or negative weight. Since the energy with respect to this problem will only depend on which edges are `cut' by joining bit variables taking opposite values and not what these values are, this problem is symmetric under inversion of all bits by construction. 

\begin{figure}
\begin{centering}
\includegraphics[width=7cm]{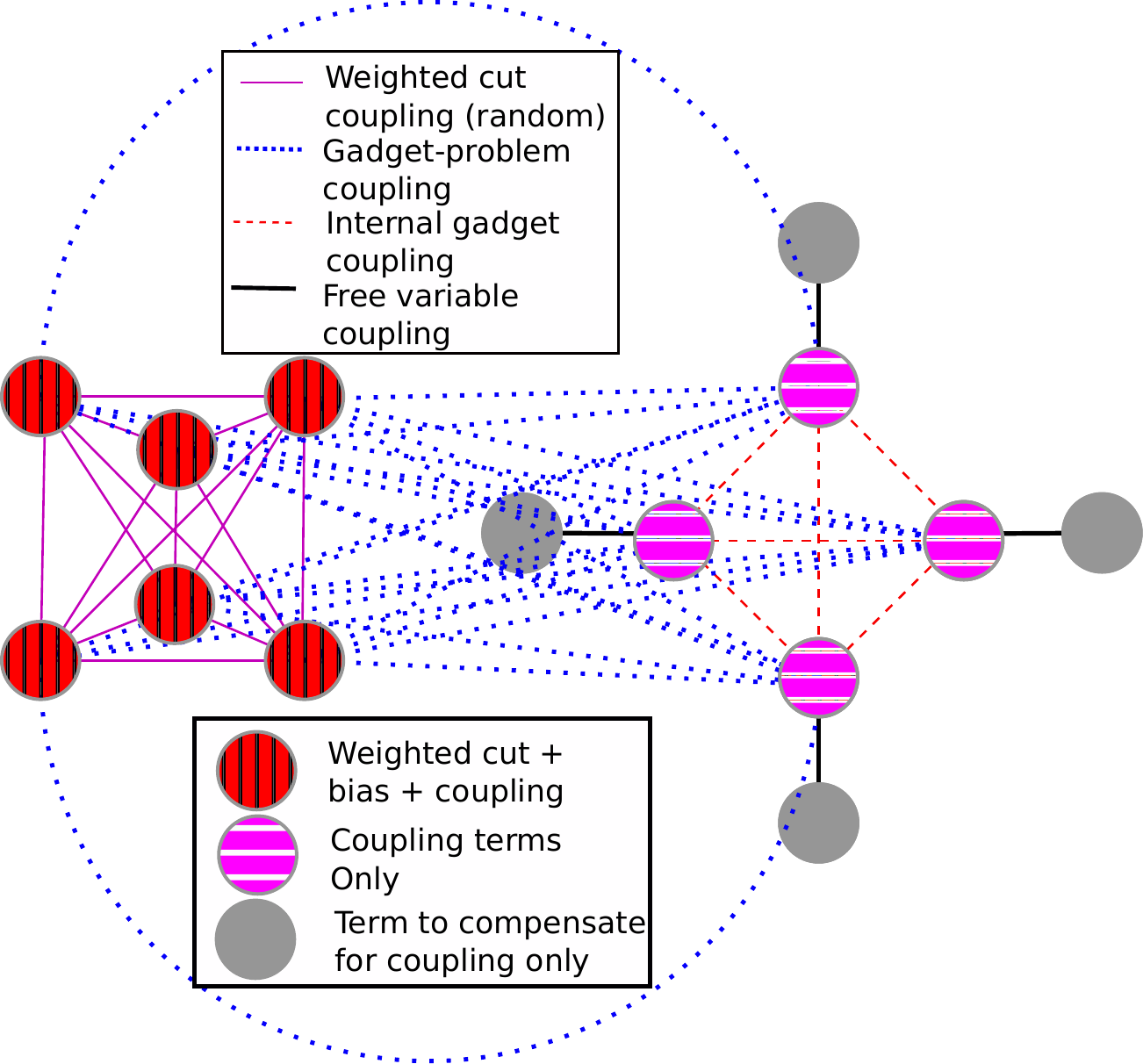}
\par
\end{centering}
\caption{Graphical representation of the diagonal and off-diagonal elements of a QUBO generated using algorithm \ref{Alg:problem_generation}. In this example $n_{\mathrm{cut}}=6$,  $n_{\mathrm{gadget}}=4$ which are the same parameters used on the example we test. Edges indicate non-zero off-diagonal QUBO elements, while nodes indicate the diagonal elements, some of which have contributions from multiple sources. The upper legend indicates the function of various off-diagonal elements, while the lower legend shows the factors which contribute to the diagonal elements corresponding to different variables.
\label{gadget}}
\end{figure}

Algorithm \ref{Alg:problem_generation} shows a procedure for generating these problems and figure \ref{gadget} shows the structure diagrammatically. In this algorithm, lines $2-10$ contain a procedure for producing a weighted maximum cut QUBO. This QUBO is then solved on line $12$ for the small instances used here. We used exhaustive search to accomplish this. Lines $13-20$ then couple this problem to a gadget which produces the false minima, generated on lines $21-32$. Finally, on lines $33-35$, a bias is introduced so that the true minima, corresponding to the bitstring defined by Sol, followed by all zeros has a lower energy than the false minima, which is comprised of the bitstring which is the logical inverse of Sol, followed by $n_{\mathrm{gadget}}$ ones, followed by an arbitrary bitstring of length $n_{\mathrm{gadget}}$. 

\begin{algorithm}
\caption{Algorithm for generating problems used to test protocols \label{Alg:problem_generation}}
\begin{algorithmic}[1]
\Procedure{Generate problem}{$n_{\mathrm{cut}}$,$n_{\mathrm{gadget}}$,$J_{\mathrm{gadget}}$,$J_{\mathrm{couple}}$,$\mathrm{Bias}$}
\State Initialize empty QUBO $Q$ of size $n_{\mathrm{cut}}+2n_{\mathrm{gadget}}$
\For{$i$ from $0$ to $n_{\mathrm{cut}}-1$}
\For{$j$ from $i+1$ to $n_{\mathrm{cut}}-1$}
\State $W \gets$ uniform random value in $[-1,1]$
\State $Q[i,j] \gets Q[i,j]+W$ 
\State $Q[j,i] \gets Q[i,j]$
\State $Q[i,i] \gets Q[i,i]-W$
\State $Q[j,j] \gets Q[j,j]-W$
\EndFor
\EndFor
\State $\mathrm{Sol} \gets$ energy minimising configuration on first $n_{\mathrm{cut}}$ variables
\For{$i$ from $0$ to $n_{\mathrm{cut}}-1$}
\For{$j$ from $n_{\mathrm{cut}}$ to $n_{\mathrm{cut}}+n_{\mathrm{gadget}}-1$}
\State $Q[i,j] \gets Q[i,j]-2\times(\mathrm{Sol}[i]-1/2)\times J_{\mathrm{couple}}/(n_{\mathrm{gadget}}\times n_{\mathrm{cut}})$ 
\State $Q[j,i] \gets Q[i,j]$
\State $Q[i,i] \gets Q[i,i]+2\times(\mathrm{Sol}[i]-1/2)\times J_{\mathrm{couple}}/(n_{\mathrm{gadget}}\times n_{\mathrm{cut}})$
\State $Q[j,j] \gets Q[j,j]+2\times(\mathrm{Sol}[i]-1/2)\times J_{\mathrm{couple}}/(n_{\mathrm{gadget}}\times n_{\mathrm{cut}})$
\EndFor
\EndFor
\For{$i$ from $n_{\mathrm{cut}}$ to $n_{\mathrm{cut}}+n_{\mathrm{gadget}}-1$}
\For{$j$ from $i+1$ to $n_{\mathrm{cut}}+n_{\mathrm{gadget}}-1$}
\State $Q[i,i] \gets Q[i,i]+2J_{\mathrm{gadget}}$ 
\State $Q[i,j] \gets Q[i,j]-J_{\mathrm{gadget}}$ 
\State $Q[j,i] \gets Q[i,j]$
\EndFor
\EndFor
\For{$i$ from $n_{\mathrm{cut}}$ to $n_{\mathrm{cut}}+n_{\mathrm{gadget}}-1$}
\State $Q[i,i+n_{\mathrm{gadget}}] \gets Q[i,i+n_{\mathrm{gadget}}]-1$ 
\State $Q[i+n_{\mathrm{gadget}},i] \gets Q[i,i+n_{\mathrm{gadget}}]$ 
\State $Q[i+n_{\mathrm{gadget}},i+n_{\mathrm{gadget}}] \gets 2$
\EndFor
\For{$i$ from $0$ to $n_{\mathrm{cut}}-1$}
\State $Q[i,i] \gets Q[i,i]+(\mathrm{Sol}[i]/2-1)\times \mathrm{Bias}/n_{\mathrm{cut}}$ 
\EndFor
\EndProcedure
\end{algorithmic}
\end{algorithm}

\section{Mixer phases in QAOA and AQA}

Before discussing the results of implementing the protocols we have discussed here, it is worth briefly explaining the effect of free variables within QAOA. While this is a subject which has been well understood in quantum annealing, see for example  \cite{Matsuda2009degeneracy,Mandra2017a,Zhang2017unfair,Boixo2013a,dickson13a,chancellor17b,Konz2019fair,Chancellor20a,Chancellor2021SearchRange}, it has been much less explored in the context of QAOA. Work has been done on applying QAOA to small problems which have small gaps and are therefore difficult for quantum annealing \cite{Zhou2020AQA}; however, this work did not include much discussion of the underlying physical source of the small gap.

\begin{figure}
\begin{centering}
\begin{subfigure}[b]{0.7\textwidth}
	\begin{centering}
	\includegraphics[width=\textwidth]{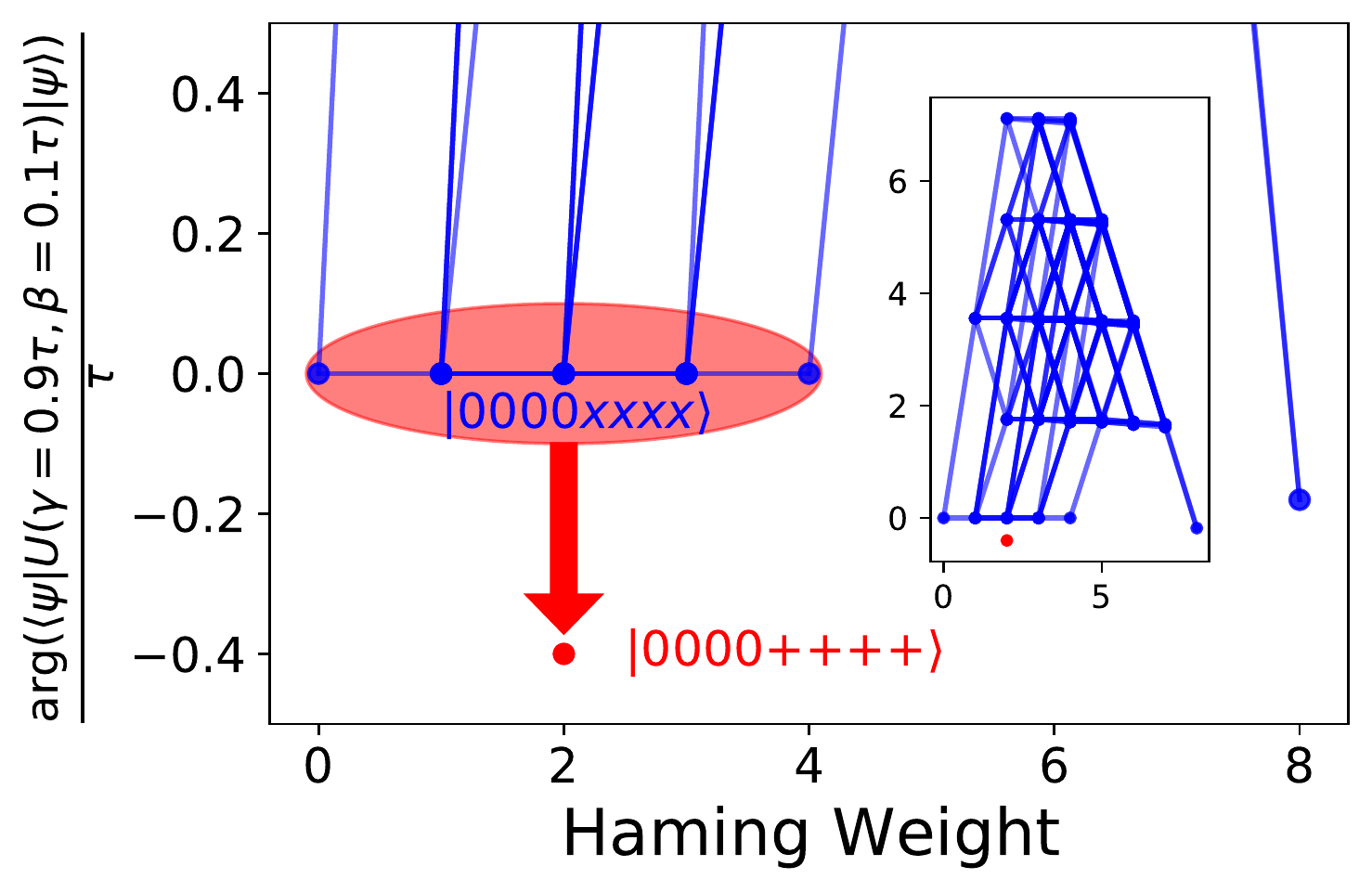}
	\par
	\end{centering}
	\caption{Hamming weight}
	\label{fig:false_min_creation}
\end{subfigure}
\begin{subfigure}[b]{0.7\textwidth}
	\begin{centering}
	\includegraphics[width=\textwidth]{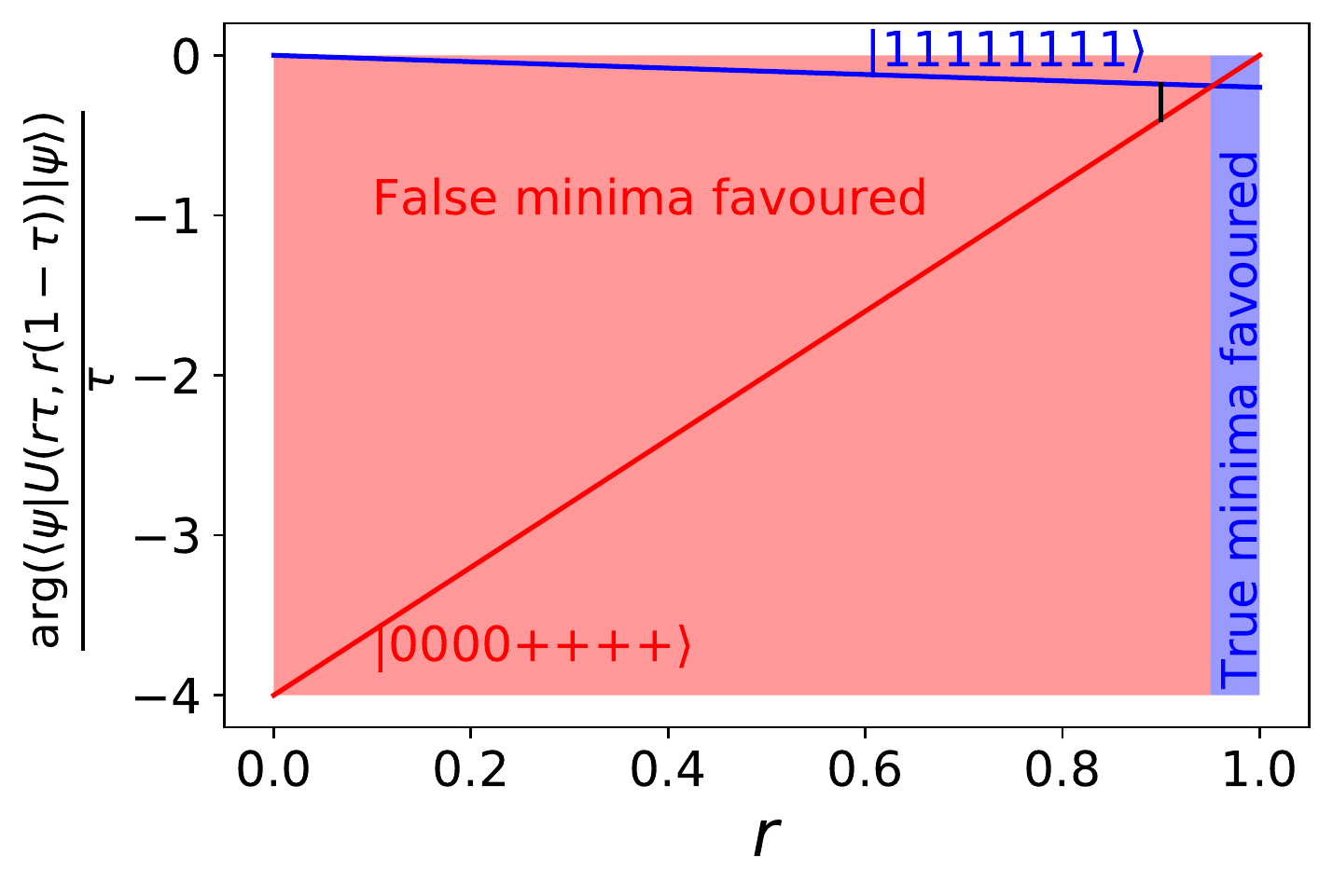}
	\par
	\end{centering}
	\caption{Relative $\gamma$ and $\beta$}
	\label{fig:phase_crossing}
\end{subfigure}
\par
\end{centering}
\caption{Phases acquired by applying AQA to the eight variable QUBO in figure \ref{fig:8_qubit}. (a) Relative phase from operating $U(\gamma,\beta)$ for different phase separator eigenstates by Hamming weight (average number of ones in the state vector); edges represent single bit flip transitions, with darker colour representing more transitions. The arrow represents the change due to phases from the mixer. The letter $x$ represents either $1$ or $0$. Inset is a zoom out of the main figure. (b) The difference between the relative phase relation for the true and false minima. Background colours represent whether the lowest phase angle is given to the false minima or the true minimum. The quantity $r=\left(1+\frac{\beta}{\gamma}\right)^{-1}$ from Eq.~\ref{eq:r_l} parameterizes the ratio between $\gamma$ and $\beta$ at different points in the protocol. The black line shows the difference seen in subfigure a.}
\end{figure}

By using engineered rather than data-mined hard problems, we can understand the source of the poor performance for AQA. Unlike in quantum annealing, there is not an obvious natural notion of spectral gap for QAOA (and if it is a course simulation, the relevance of the annealing Hamiltonian spectral gap to AQA will be lower\footnote{For fast anneals the minimum spectral gap is also likely to be less relevant and mechanisms like those described in \cite{Callison21a,crosson2020prospects} are likely to be more important.}). We instead base our intuition on the phases a state acquires when the phase separator and mixer are performed sequentially for different values of $\gamma$ and $\beta$. We therefore define:
\begin{equation}
    \phi(\ket{\psi},\gamma,\beta)=\arg(\sandwich{\psi}{U(\gamma,\beta)}{\psi}) \label{eq:phase_def}
\end{equation}
where $U(\gamma,\beta)$ is the unitary defined by a single layer of the QAOA circuit, equation \ref{eq:traditional_layer}, and $\ket{\psi}$ is the state it is applied to. For the purposes of this analysis, when we define the $\arg$ operation which recovers the phase, we consider a branch cut at $\pi$ and further limit our analysis to cases where $\gamma$ and $\beta$ such that $|\phi|\ll\pi$. Therefore, we do not have to worry about large positive and large negative rotations appearing to be identical. Since we are interested in solving a problem represented by the phase separator, it is natural to consider eigenstates of the phase separator which are the $\ket{\psi}$ states.  All computational basis states will be eigenstates of the phase separator, since it is diagonal in the $z$ basis. However, free variables, which give the same phase contribution for $\ket{0}$ and $\ket{1}$ configuration, can be in any configuration, and the state will still be an eigenstate of the phase separator. If the qubits corresponding to these variables take the $\ket{+}$ configuration, then the state will acquire an additional $-\beta$ of phase rotation from the mixer operation for each qubit in this configuration. As an example, consider the QUBO given in figure \ref{fig:8_qubit}. Figure  \ref{fig:false_min_creation} illustrates this process, with an eigenstate of $U(\gamma,\beta)$ with support over $16$ classical basis states. 

While we have not done it here, it would be possible to instead consider equation \ref{eq:phase_def} but setting the states $\ket{\psi}$ to be the eigenstates of $U(\gamma,\beta)$. In the limit of small $\gamma$ and $\beta$, the operation $U(\gamma,\beta)$ would become an approximation of the instantaneous action of a quantum annealing Hamiltonian, and this would therefore reduce to the instantaneous energy eigenstates which are often used in analysis within that setting.

\section{Mixer phase effects on an IonQ QPU}

Before numerically examining strategies to mitigate against the phase effects discussed in the previous section, it is worth demonstrating that signatures of these effects can be seen on real devices. While the simulations performed later in this work will all be based on the ideal setting, real devices have noise and other imperfections. It is therefore useful to demonstrate that the effects here are relevant on such devices. For these experiments, we use AQA, as defined before, on an IonQ Harmony processor with a simple three qubit QUBO, $Q_{0,1}=Q_{1,0}=-1$, $Q_{1,2}=Q_{2,1}=0.5$, $Q_{0,0}=Q_{1,1}=1$ and all other elements zero. The ground state manifold of this QUBO consists of $\{\ket{000},\ket{001},\ket{110}\}=\{\ket{00+},\ket{00-},\ket{110}\}$. As a result, the $\ket{00+}$ state will be able to acquire additional phases which will mimic the effect of a lower energy while the additional phase contributions to $\ket{00-}$ will mimic a higher energy. We therefore expect that the state $\ket{00+}$ will be over represented in our samples relative to the other two, while $\ket{00-}$ will be under represented. Our experimental methods are explained in section \ref{sec:exp_meth}.

\begin{figure*}[htp!]
\begin{subfigure}[b]{0.475\textwidth}
	\begin{centering}
	\includegraphics[width=\textwidth]{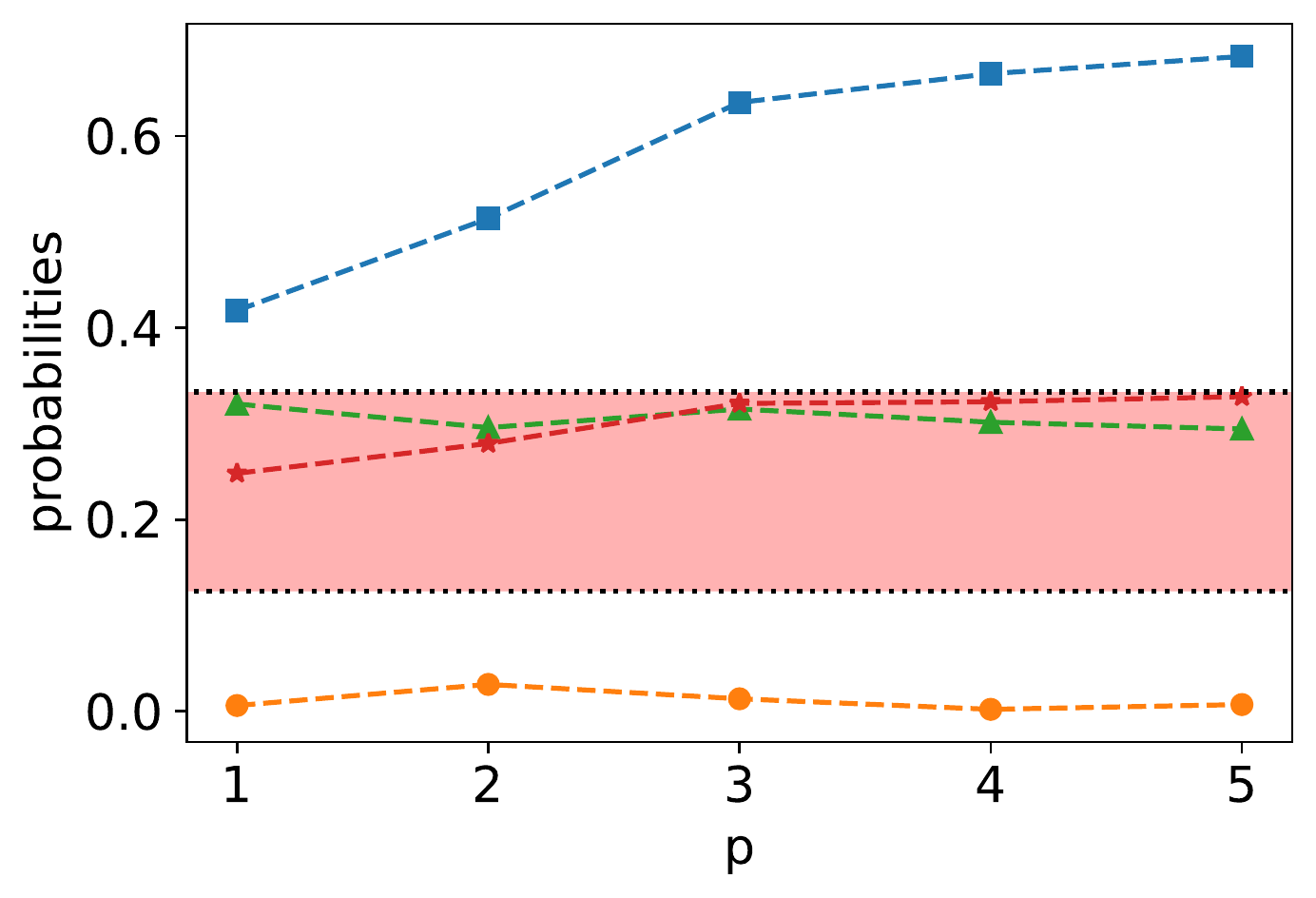}
	\par
	\end{centering}
	\caption{Ideal Simulation}
	\label{fig:simple_cirqsim_no_sup}
\end{subfigure}
\begin{subfigure}[b]{0.475\textwidth}
	\begin{centering}
	\includegraphics[width=\textwidth]{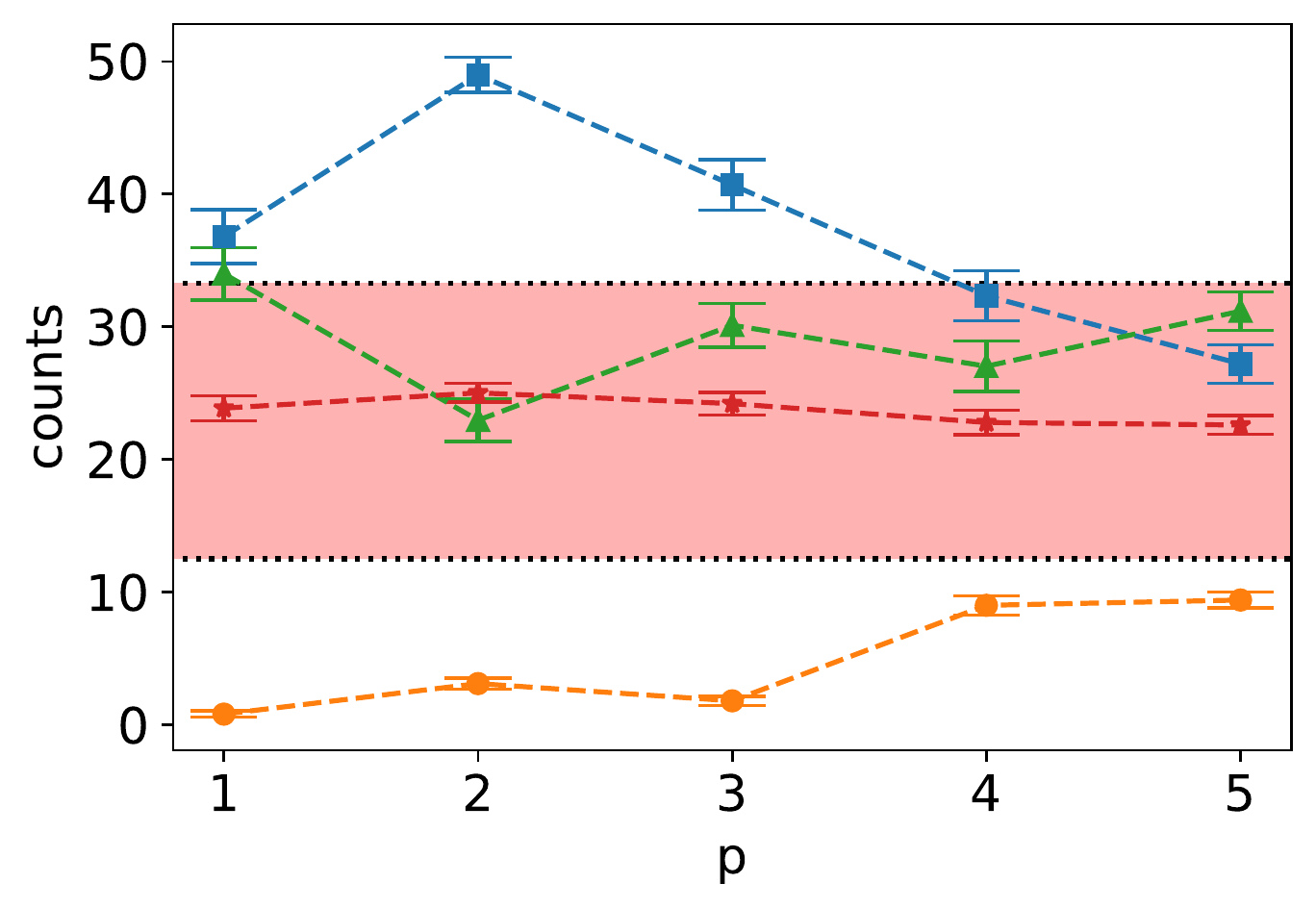}
	\par
	\end{centering}
	\caption{IonQ Harmony}
	\label{fig:simple_harmony_no_sup}
\end{subfigure}
\caption{Probability of different states for AQA solving a simple three bit QUBO $Q_{0,1}=Q_{1,0}=-1$, $Q_{1,2}=Q_{2,1}=0.5$, $Q_{0,0}=Q_{1,1}=1$ (all other elements $0$) with $\tau=\frac{3}{4}\pi$. Squares (blue) represent counts/probability for the $\ket{00{+}}$ state; triangles (green) represent an upper bound on the counts/probability for $\ket{110}$; circles (orange) represent counts/probability for the $\ket{00{-}}$ state; and stars (red) represent the mean of the three. Harmony was run ten times with $100$ samples each; error bars represent standard error. The shaded areas are a guide showing the range of average values between completely random guessing and perfect recovery of states within the ground state manifold. For both plots $\zeta_2=1$.}
\end{figure*}

An experimental difficulty is that unless we compile a complex gate such as a controlled Hadamard, we cannot simultaneously measure the rate at which $\{\ket{00+},\ket{00-},\ket{110}\}$ are all observed. To circumvent this issue, we instead use an upper bound for what the maximum probability associated with $\ket{110}$ could be, based on the number of observations of $\ket{11+}$ and $\ket{11-}$. The probability of $\ket{110}$ would be maximised if the (unknown) phase between $\ket{11+}$ and $\ket{11-}$ were positive, so we calculate the probability under this assumption. Firstly, we perform an idealised numerical simulation to verify that under ideal circumstances we would indeed observe a relative enhancement of $\ket{00+}$ and suppression of $\ket{00-}$. As figure \ref{fig:simple_cirqsim_no_sup} shows, the effect of phases from the mixer is visible even in this simple three-qubit system. The most obvious manifestation of this effect is the large enhancement of $\ket{00+}$ at the expense of $\ket{00-}$. There is also a more subtle effect visible for $p>2$ in the slight suppression of the $\ket{110}$ state relative to the mean probability. The reason that the second of these effects is slight is likely that it is relatively easy to flip only two bits, even with a relatively weak mixer, therefore transitions between the $\ket{110}$ state and the $\{\ket{00+}, \ket{00-}\}$ manifold will be significant even late in the protocol where the accumulated phases in these states are slight. It is also worth noting that for $p\ge 3$, the ideal version is able to find the ground state manifold nearly perfectly.

For the experimental test results depicted in \ref{fig:simple_harmony_no_sup}, we see that for $p=1$ and $p=2$, the theoretically modelled trend of increasing probability of $\ket{00+}$ with $p$ is experimentally reproduced, but then reverses for larger values. This is likely due to noise within the device. However, at all $p$ values tested, there is a significant enhancement of $\ket{00+}$ relative to $\ket{00-}$ although this enhancement decreases with $p$, and for $p=1,2,3$ there is a significant enhancement of $\ket{00+}$ over $\ket{110}$. Both of these indicate the presence of mixer phase effects and that noise starts to mask the effects at larger values of $p$. The strongest signature we observe is reduction of the $\ket{00-}$ state probability, which will be useful when measuring the effect of fluctuation suppression schemes. While the ideal simulation shows a trend of continued low probability of observing $\ket{00-}$ as $p$ is increased, this probability generally increases with $p$ on the real device. We also remark that in the case of the real device, the average probability to be found in one of these states is constant within statistical error over the $p$ values tested and lies between what would be expected from completely random guessing and what would be expected for complete success in finding the ground state manifold.

\begin{figure*}[htp!]
\begin{subfigure}[b]{0.475\textwidth}
	\begin{centering}
	\includegraphics[width=\textwidth]{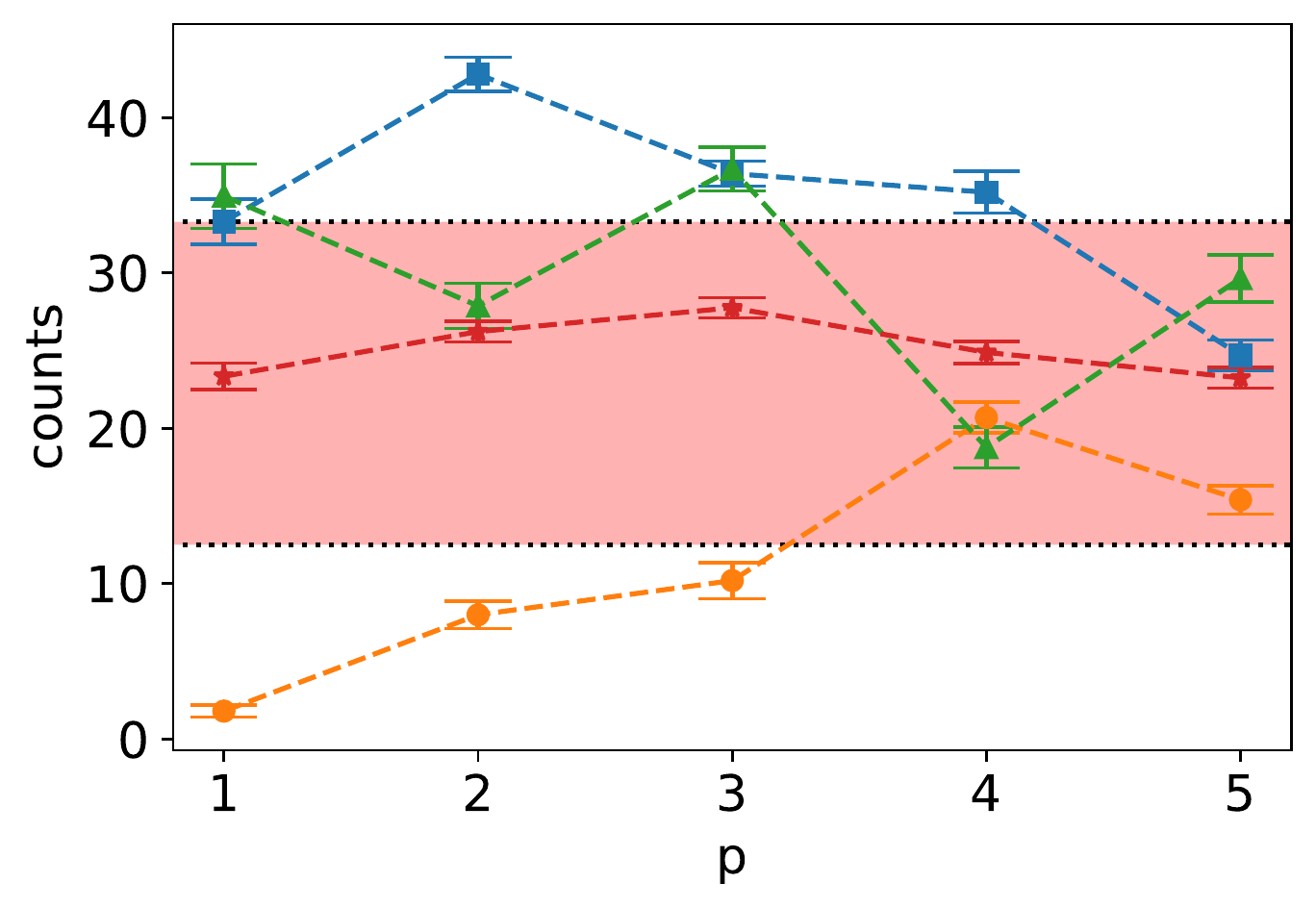}
	\par
	\end{centering}
	\caption{$\zeta_2=\frac{1}{2}$}
	\label{fig:simple_harmony_sup}
\end{subfigure}
\begin{subfigure}[b]{0.475\textwidth}
	\begin{centering}
	\includegraphics[width=\textwidth]{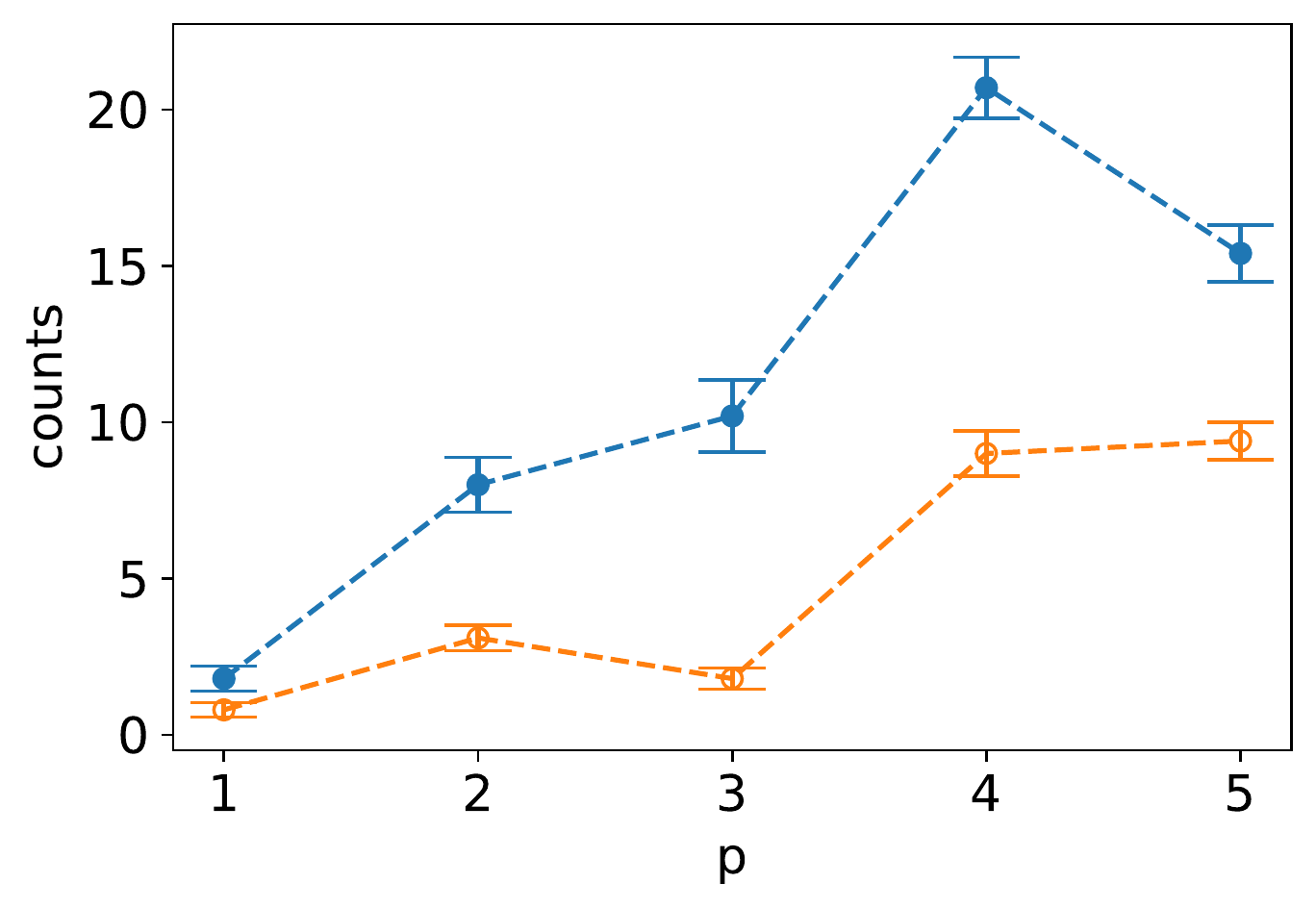}
	\par
	\end{centering}
	\caption{$\ket{00{-}}$ count comparison}
	\label{fig:zzm_compare}
\end{subfigure}
\begin{subfigure}[b]{0.475\textwidth}
	\begin{centering}
	\includegraphics[width=\textwidth]{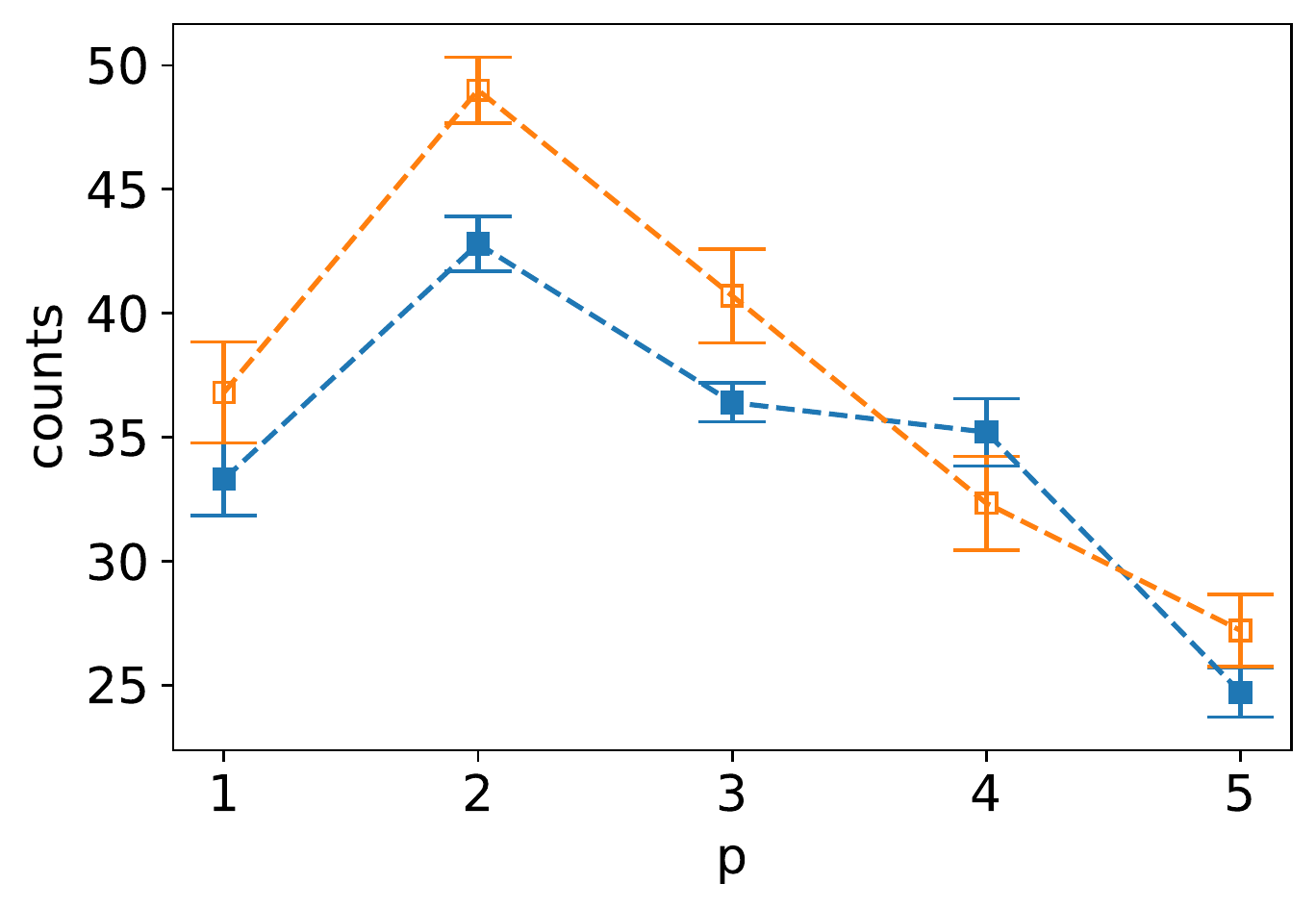}
	\par
	\end{centering}
	\caption{$\ket{00{+}}$ count comparison}
	\label{fig:zzp_compare}
\end{subfigure}
\begin{subfigure}[b]{0.475\textwidth}
	\begin{centering}
	\includegraphics[width=\textwidth]{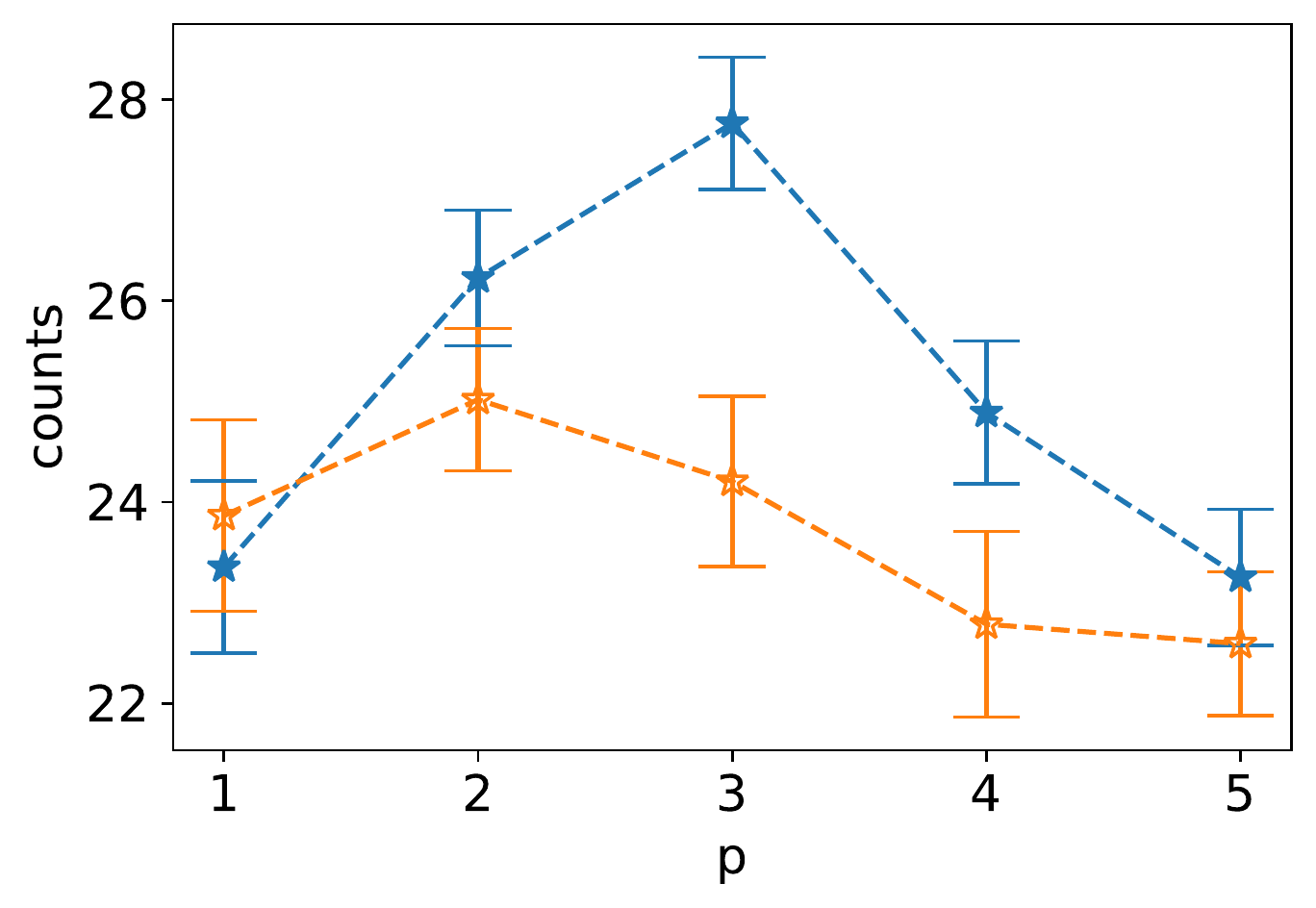}
	\par
	\end{centering}
	\caption{mean count comparison}
	\label{fig:mean_compare}
\end{subfigure}
\caption{(a) Probabilities from the same configuration as \ref{fig:simple_harmony_no_sup}, but suppressing fluctuations by setting $\zeta_2=\frac{1}{2}$ as opposed to the default $\zeta_2=1$. Squares (blue) represent counts for the $\ket{00{+}}$ state; triangles (green) represent an upper bound on the counts for $\ket{110}$; circles (orange) represent counts for the $\ket{00{-}}$ state; and stars (red) represent the mean of the three. For other figures, the symbols, but not the colors from (a) are preserved. (b) Comparison of  counts for the $\ket{00{-}}$ state. (c) Comparison of  counts for the $\ket{00{+}}$ state. (d) Comparison of mean counts. For (a,b,c), filled symbols (blue) represent $\zeta_2=\frac{1}{2}$ (suppressed), and unfilled symbols (orange) represent $\zeta_2=1$ (not suppressed). Error bars represent standard error.\label{fig:simple_sup_compare}}

\end{figure*}

Next, we want to experimentally verify the suppression strategy. In other words, we want to show that selectively reducing the driver rotation angle on qubit $2$ can indeed reduce the effects we attribute to the mixer phases. To experimentally test this, we set $\zeta_2=\frac{1}{2}$ (as opposed to the previous test where $\zeta_2=\zeta_1=\zeta_0=1$), effectively halving the mixing angle on the fluctuating qubit at all points in the circuit. The first thing we notice by comparing \ref{fig:simple_harmony_sup} with \ref{fig:simple_harmony_no_sup} is that the probability to be found in $\ket{00{-}}$ is strongly increased by reducing the phase angle rotation for $p>1$. This difference can be seen more clearly by plotting the curves together in figure \ref{fig:zzm_compare}, and this difference is much larger than the error bars for all $p>1$. This is consistent with an interference mechanism due to phases arising from the mixer, as these phases will be larger with $\zeta_2=1$ and will effectively behave as if this state has a higher energy. The lack of significant difference for $p=1$ is also consistent with this picture, since the mixer is only applied once, and while phases will be accumulated due to this mixer term, there is no opportunity for interference after it is applied.

If, on the other hand, we examine the counts of the $\ket{00{+}}$ state as is done in figure \ref{fig:zzp_compare}, we see that there is a significant enhancement of the probability of this state for the runs using $\zeta_2=1$ for $p=2$, a marginally significant enhancement at $p=3$, and barely outside of error bars for other values (recall that error bars are standard error, so points being slightly more than the bars away from each other does not indicate a statistically strong result). This effect is in the direction we expect due to mixer phases and is likely not visible at higher $p$ due to the fact that this quantity appears more sensitive to experimental noise. For $p=1$, an effect is not expected because, while the phase is applied, there is no chance for subsequent interference.

Finally, we should examine the total probability to be found in the ground state manifold. While there is not necessarily an obvious trend which should be intuitively expected here due to the mixer phases, the comparison is still worth making. This comparison allows us to rule out, for example, the results in figure \ref{fig:zzm_compare} being due to a difference in ability to find any ground state rather than an imbalance between them. We find in figure \ref{fig:mean_compare} that results here are only far outside of error bars for $p=3$ (and moderately so for $p=2,4$), with the $\zeta_2=
\frac{1}{2}$ simulation finding the ground state manifold more often. It is worth noting that while statistically significant (at least for $p=3$), even this difference is significantly smaller than those seen in figure \ref{fig:zzm_compare} (note that the y scale is not the same across these figures). We can therefore conclude that the results in \ref{fig:zzm_compare} are indeed primarily due to the distribution between states within the ground state manifold, not the ability to find the manifold at all. 

While there are clear arguments for which states within the ground state manifold should be most strongly represented based on the mixer phase argument, and further that suppression of $\zeta_2$ should reduce these differences, there is no clear intuition we could find as to what the effect on the overall probability of finding the ground state would be. A full understanding would likely require examining a model including noise, since the ground states can easily be found in the noise-free system. Since these effects are not relevant to the main conclusion of this study, we have elected not to investigate this behavior, although it may be an interesting area of investigation in the future.

\section{Numerical Results}

\subsection{Original problem from \cite{dickson13a}\label{sub:Dickson_orig}}

\begin{figure*}[htp!]
\begin{subfigure}[b]{0.475\textwidth}
	\begin{centering}
	\includegraphics[width=\textwidth]{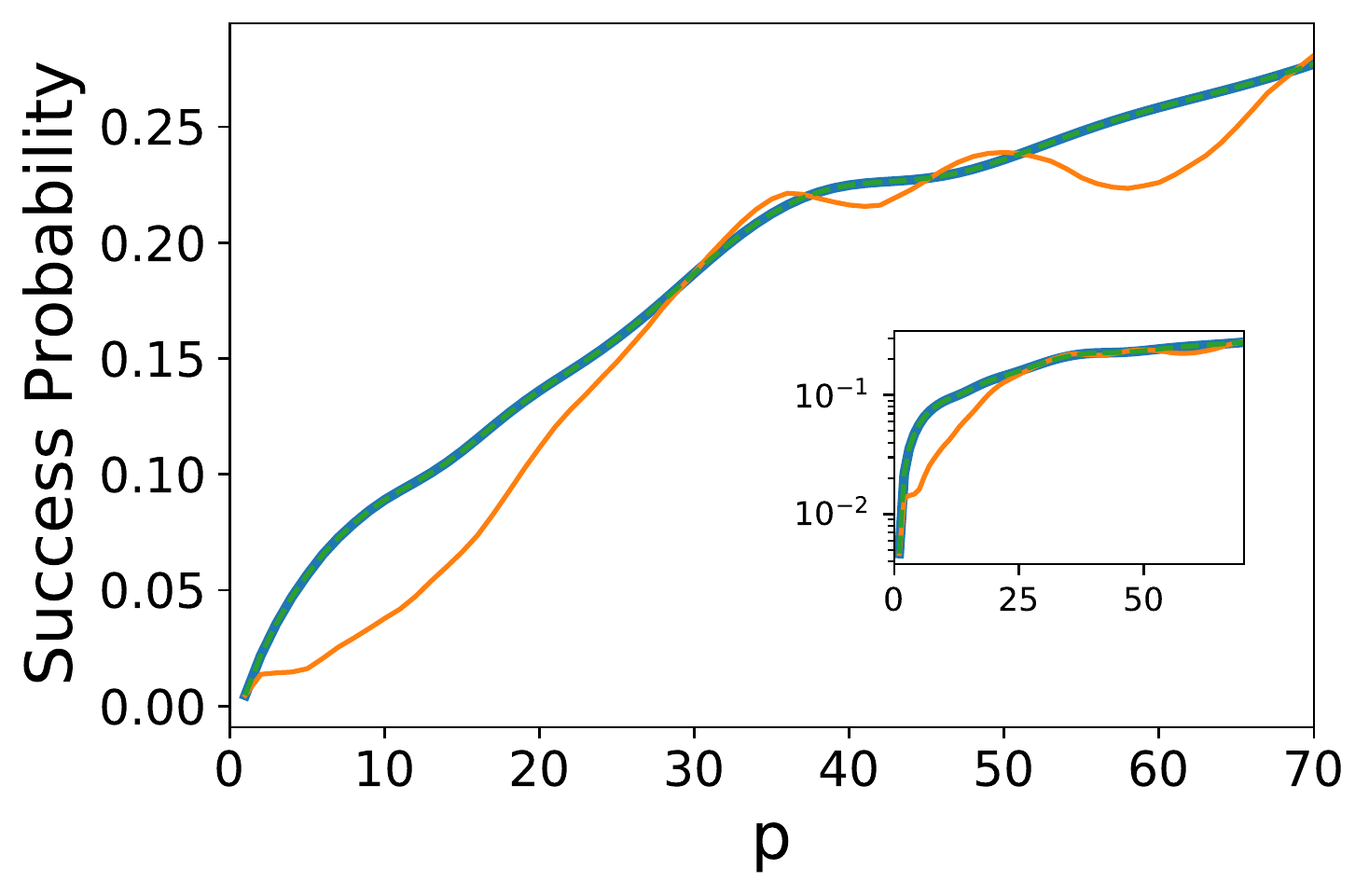}
	\par
	\end{centering}
	\caption{Success probabilities}
	\label{fig:TA_1_p_1_4_success}
\end{subfigure}
\begin{subfigure}[b]{0.475\textwidth}
	\begin{centering}
	\includegraphics[width=\textwidth]{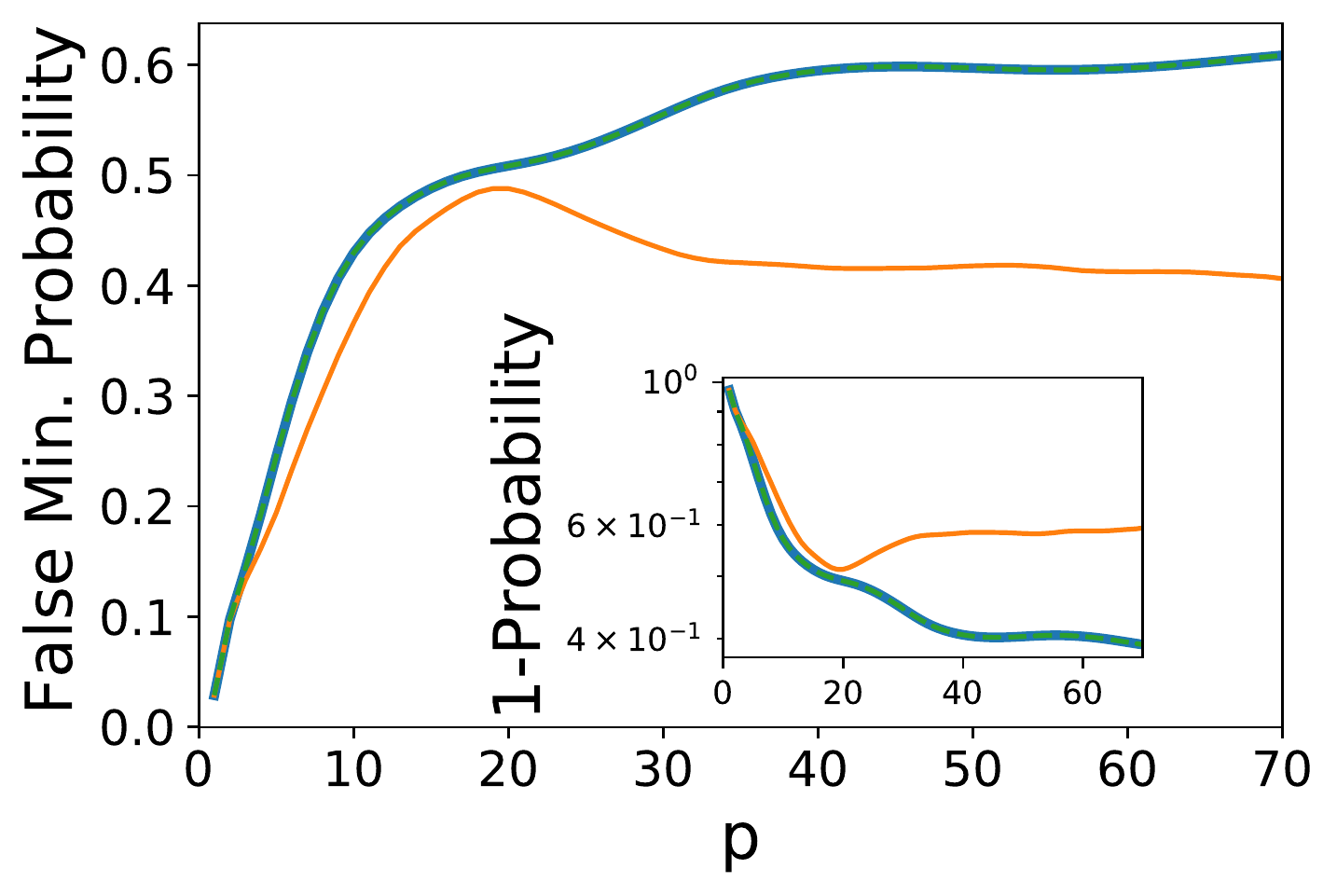}
	\par
	\end{centering}
	\caption{False minimum probabilities}
	\label{fig:TA_1_p_1_4_false}
\end{subfigure}
\caption{Success and false minimum probability for a modified the problem used in \cite{dickson13a}  with $\tau(p)=\frac{\pi}{2p^{0.25}}$. Blue corresponds to an unmodified mixer, while gold corresponds to the suppressed mixer and green to a thresholded mixer. Because the unmodified and thresholded curves completely overlap we have visually indicated this by making the thresholded curve dashed and increasing the width of the unmodified curve. The inset is the same as the main plot but on a logarithmic scale. }
\end{figure*}

To test the general concept that mixer adjustments based on the diagonal elements of the Fubini-Study metric will increase the chance of finding the true minima, we start with a problem statement corresponding exactly to the one used in \cite{dickson13a} with the structure depicted in figure \ref{fig:16_qubit}. Since it behaves like quantum annealing in the limit of large $p$, we elect to examine a schedule where the individual angles are decreased $\tau(p)=\frac{\pi}{2p^{0.25}}$. For the linear protocol, none of the diagonal elements of the Fubini-Study metric ever fall below $\Theta=0.2$ for the values of $p$ we examine.  Figure \ref{fig:TA_1_p_1_4_success} shows that for both the suppressed and unmodified mixer, there is a general trend of increasing success probability with increasing $p$. Because the diagonal metric elements never fall below the threshold, the thresholded and unmodified protocols are equivalent for this problem.  To understand the differences between the suppressed and unmodified behavior, we can further examine the probability to be in a false minima from figure \ref{fig:TA_1_p_1_4_false}. From this figure, we see a generally increasing trend for the unmodified mixer, but one which is far from unity for the $p$ values we examine. We do see that the suppressed mixer reverses the trend at $p\gtrsim15$, suggesting that the suppression is acting as desired.

\subsection{Modified version\label{sub:Dickson_mod}}

\begin{figure*}[htp!]
\begin{subfigure}[b]{0.475\textwidth}
	\begin{centering}
	\includegraphics[width=\textwidth]{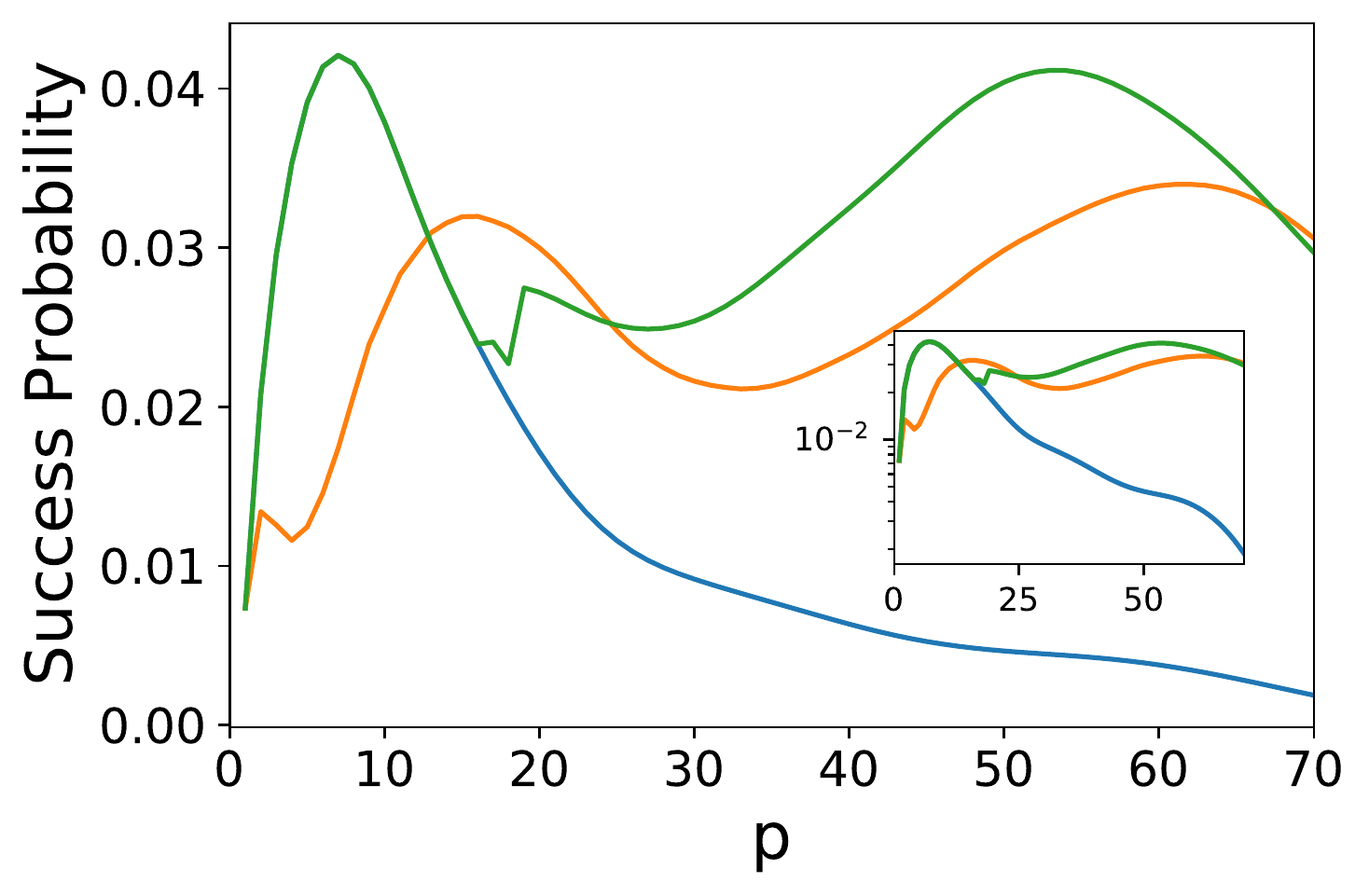}
	\par
	\end{centering}
	\caption{Success probabilities}
	\label{fig:TA_0p2_p_1_4_success}
\end{subfigure}
\begin{subfigure}[b]{0.475\textwidth}
	\begin{centering}
	\includegraphics[width=\textwidth]{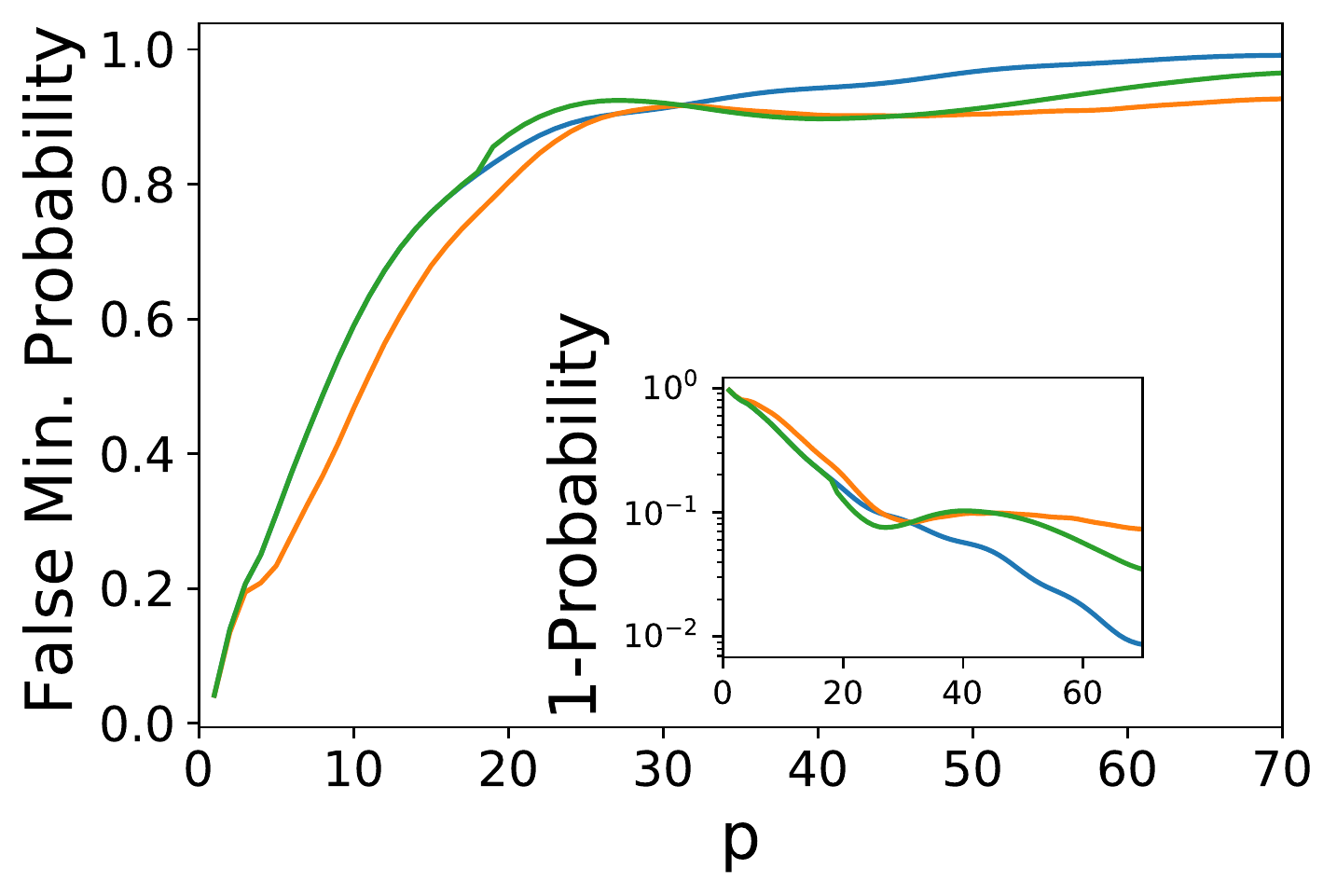}
	\par
	\end{centering}
	\caption{False minimum probabilities}
	\label{fig:TA_0p2_p_1_4_false}
\end{subfigure}
\caption{Success (a) and false minimum (b) probability for the problem used in \cite{dickson13a}  with a factor of $5$ smaller separation between the false minimum and true ground state and $\tau(p)=\frac{\pi}{2p^{0.25}}$. Blue corresponds to an unmodified mixer, while gold corresponds to the suppressed mixer and green to a thresholded mixer. The inset is the same as the main plot but on a logarithmic scale. }
\end{figure*}

\begin{figure}
\begin{centering}
\includegraphics[width=7cm]{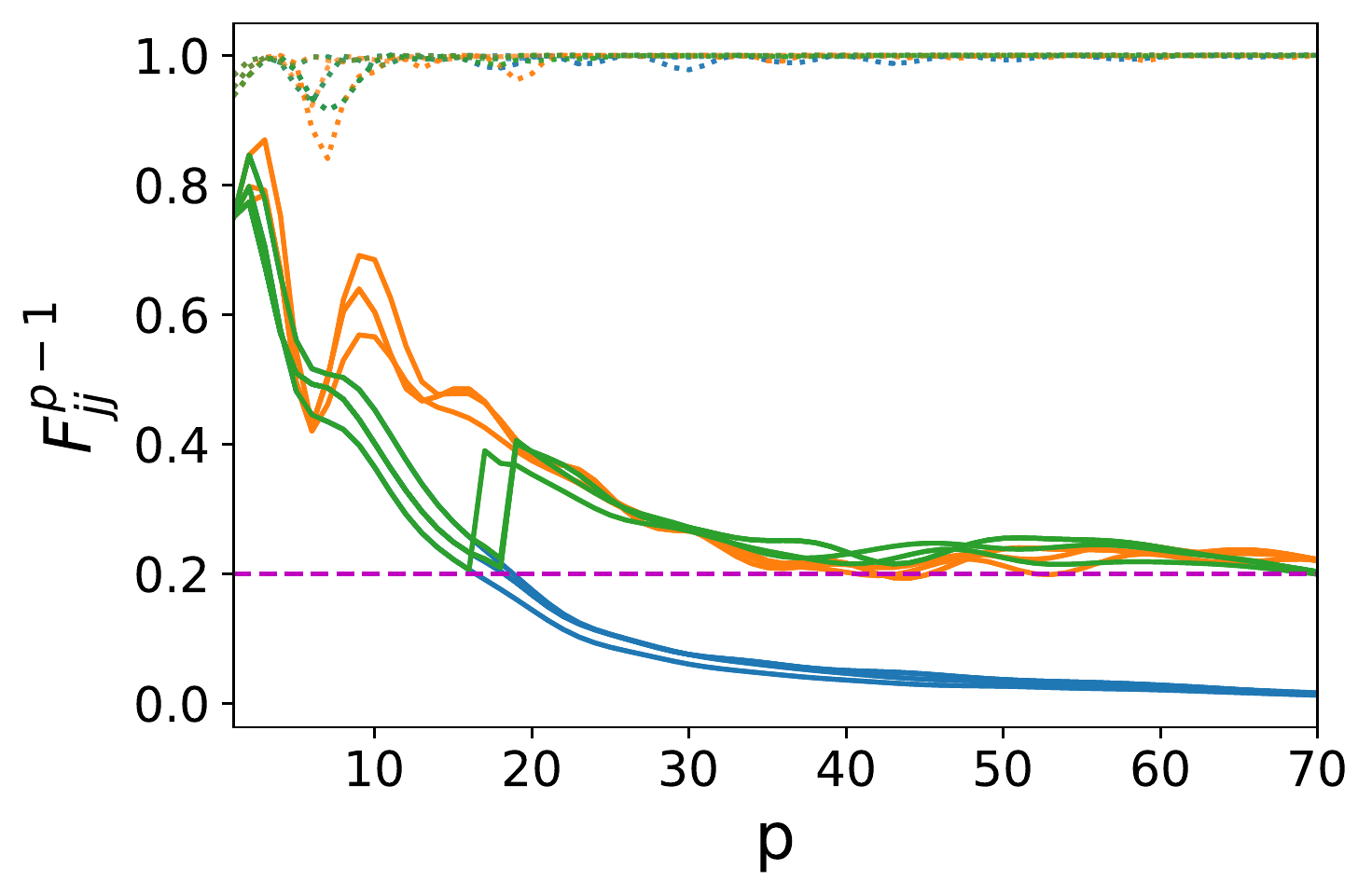}
\par
\end{centering}
\caption{Diagonal elements of the Fubini Study metric for the problem used in \cite{dickson13a} with a factor of $5$ smaller separation between the false minimum and true ground state and $\tau(p)=\frac{\pi}{2p^{0.25}}$. Blue corresponds to an unmodified mixer, while gold corresponds to the suppressed mixer and green to a thresholded mixer. Curves corresponding to variables which do not become free late in the anneal are shown as fainter dotted lines, while those corresponding to variables which are free in the false minimum are shown as solid lines. The dashed line indicates the threshold of 0.2. Each curve corresponds to a different metric element. \label{fig:TA_0p2_p_1_4_QGT_final}}
\end{figure}

To enhance the effects of the false minima, we reduce the energy difference between the true and false minima by a factor of $5$. This makes it harder to escape the false minima, and in turn, makes it so that the diagonal metric terms go below the threshold, leading to a difference between thresholded and unmodified AQA. Figure  \ref{fig:TA_0p2_p_1_4_success} illustrates a clear difference between suppressed and thresholded mixers and an unmodified mixer for this problem. For the unmodified version, a feature known as a ``diabatic bump'' \cite{Crosson14a,Muthukrishnan2016tunneling,Hormozi2017nonstoquatic,Albash2018adiabatic,Zhou2020AQA}, which has been previously seen in other studies of annealing like protocols, is clearly visible. While the success probabilities under the two protocols using modified drivers both tend to oscillate with increasing $p$, the success probability corresponding to the unmodified mixer drops dramatically as $p$ is increased. As can be seen in Figure \ref{fig:TA_0p2_p_1_4_false}, this problem also shows a high probability for the false minimum to be found, and at large $p$, this probability is higher for the unmodified mixer. A moderate amount of instability can be seen from figure \ref{fig:TA_0p2_p_1_4_QGT_final}, which shows that the individual qubits do not cross the threshold at the same time, and that there is a relatively large jump in the value when the unmodified version crosses the threshold.

This example problem, however, does not directly demonstrate an advantage from suppressing fluctuations since peak performance is reached at relatively low $p$, before there is a strong tendency for the dynamics to get stuck in a local minimum and before the threshold for suppression on any of the qubits is reached. We can see that the suppressed and thresholded mixers are fulfilling their desired role of reducing fluctuations which distort the search of the energy landscape, and the final values of the diagonal elements of the Fubini-Study metric corresponding to the variables which are free in the local minimum can be significantly increased using these techniques.

\subsection{Specialised QUBO\label{sub:spec_qubo}}

\begin{figure*}[htp!]
\begin{subfigure}[b]{0.475\textwidth}
	\begin{centering}
	\includegraphics[width=\textwidth]{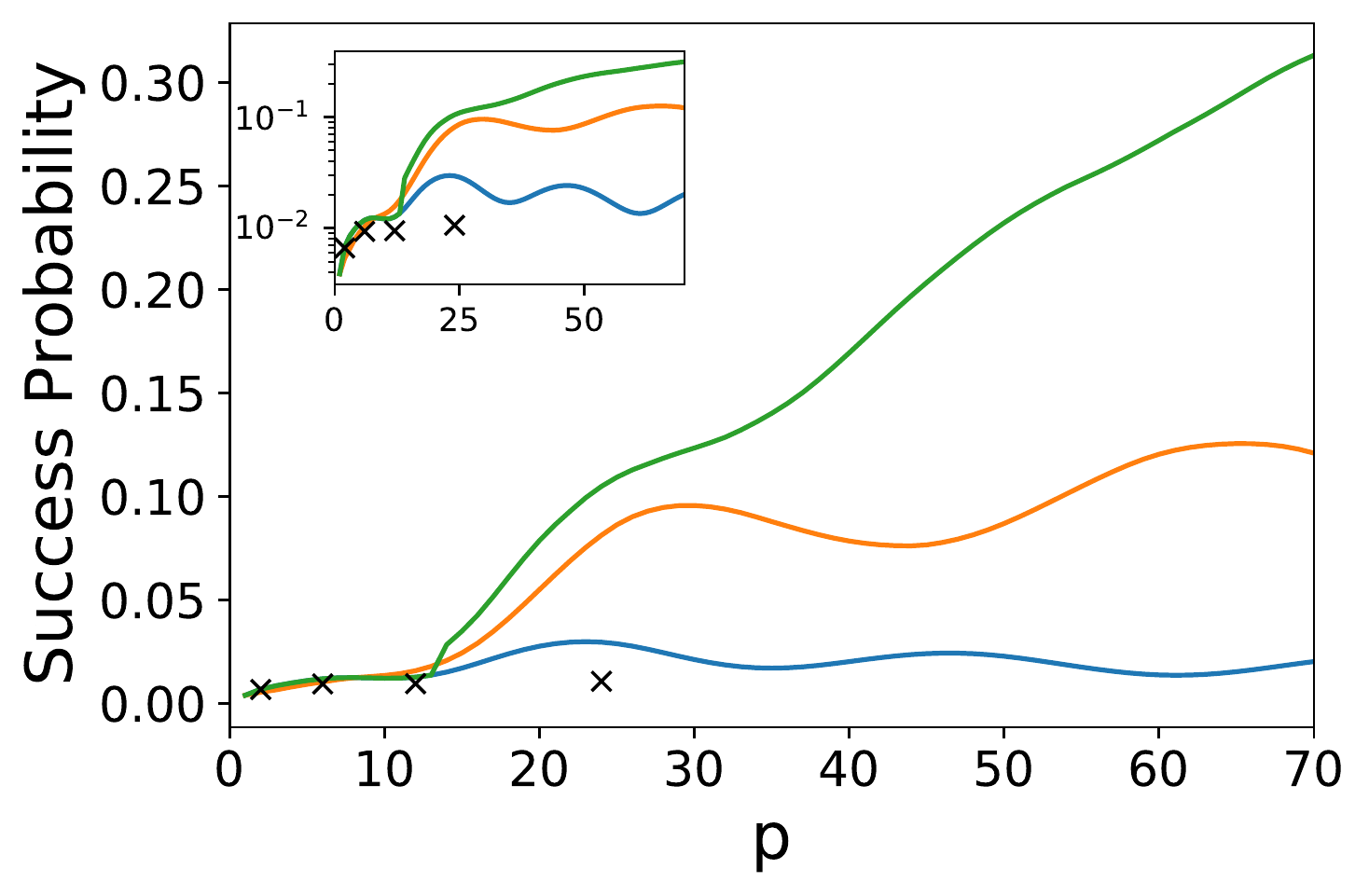}
	\par
	\end{centering}
	\caption{Success probabilities}
	\label{fig:gad_1_p_1_4_success}
\end{subfigure}
\begin{subfigure}[b]{0.475\textwidth}
	\begin{centering}
	\includegraphics[width=\textwidth]{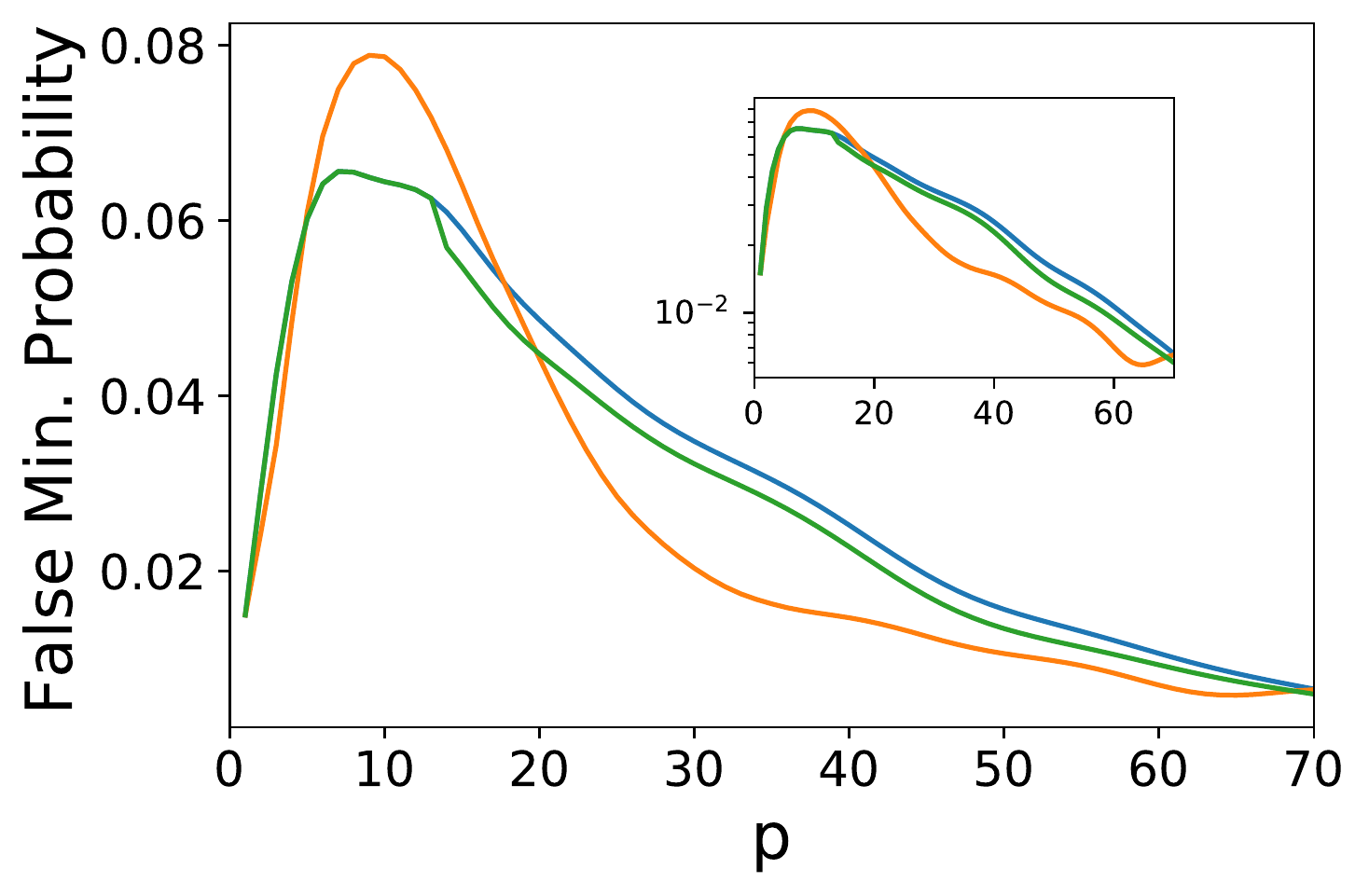}
	\par
	\end{centering}
	\caption{False minimum probabilities}
	\label{fig:gad_1_p_1_4_false}
\end{subfigure}
\caption{Success and false minimum probability for Pauli X operators at the end of AQA solving a QUBO created using algorithm \ref{Alg:problem_generation} using $n_\mathrm{cut}=6$, $n_\mathrm{gadget}=4$, $J_{gadget}=0.25$, $J_{\mathrm{couple}}=0.5$, $\mathrm{Bias}=1.5$ and $\tau(p)=\frac{\pi}{2p^{0.25}}$. Blue corresponds to an unmodified mixer, while gold corresponds to the suppressed mixer and green to a thresholded mixer. The crosses on the success probability plot represent the results of QAOA, using the methods described in section \ref{sec:num_meth}, is the average value of the best performing iteration of 100 runs QAOA optimisation with each run consisting of 300 iterations. Convergence of mean value is demonstrated in appendix I, figure \ref{fig:QAOA_converge}.  The inset is the same as the main plot but on a logarithmic scale. The specific QUBO used in these plots can be found in table \ref{tab:spec_QUBO} in appendix II. \label{fig:gad_1_p_1_4}}
\end{figure*}

To find a regime where the suppressed and thresholded mixers can directly be beneficial, we consider an instance of a specialised QUBO created using algorithm \ref{Alg:problem_generation}. These examples are more realistic to the real world because the weighted maximum cut elements of the problem statement will create rough features in the energy landscape similar to what would be encountered in a real-world optimisation problem, while also having an engineered false minima feature. We examine a version with $n_\mathrm{cut}=6$, $n_\mathrm{gadget}=4$, $J_{gadget}=0.25$, $J_{\mathrm{couple}}=0.5$, $\mathrm{Bias}=1.5$. Figure \ref{fig:gad_1_p_1_4_success} shows the success probability from applying QAOA with the three different mixer strategies to this QUBO. The thresholded and suppressed methods both significantly outperform an unmodified mixer, with the thresholded technique performing the best, and a general upward trend in success probability with increasing $p$. Furthermore, we see that using an optimised choice of $\gamma$ and $\beta$ does not in itself alleviate the problems caused by the pathological features of this problem statement. We see in figure \ref{fig:gad_1_p_1_4_success} that at $p=24$, a more traditional QAOA strategy actually performs somewhat worse than traditional AQA with a linear schedule. This makes sense because the optimisation strategy optimises against energy expectation, rather than the actual success probability (which would not be possible to measure without knowing the solution), and the false minimum also has a low (although not quite optimal) energy expectation.

\begin{figure}
\begin{centering}
\includegraphics[width=7cm]{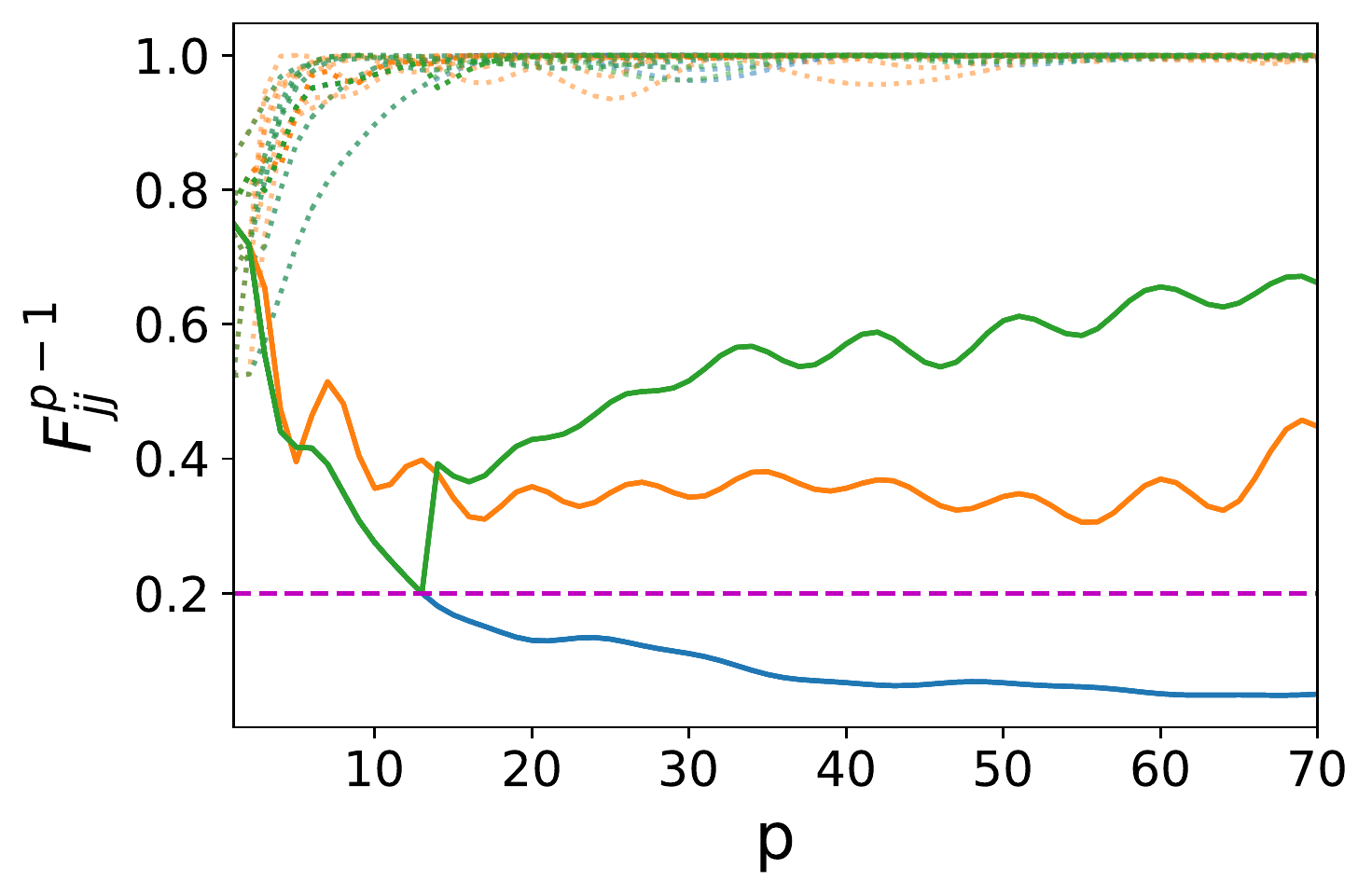}
\par
\end{centering}
\caption{Diagonal elements of the Fubini Study-metric for Pauli X operators at the end of QAOA solving a QUBO created using algorithm \ref{Alg:problem_generation} using $n_\mathrm{cut}=6$, $n_\mathrm{gadget}=4$, $J_{gadget}=0.25$, $J_{\mathrm{couple}}=0.5$, $\mathrm{Bias}=1.5$ and $\tau(p)=\frac{\pi}{2p^{0.25}}$. Blue corresponds to an unmodified mixer, while gold corresponds to the suppressed mixer and green to a thresholded mixer. The dashed line indicates the threshold of 0.2. Each curve corresponds to a different metric element. The specific QUBO used in these plots can be found in table \ref{tab:spec_QUBO} in appendix II. \label{fig:gad_1_p_1_4_QGT_final}}
\end{figure}

We find, however, that unlike in the case of the modified version of the problem statement from \cite{dickson13a}, the cases where the true minimum is not found are not dominated by the false minimum. Figure \ref{fig:gad_1_p_1_4_false} shows that not only is the probability of being in the false minimum relatively small, it starts dropping off beyond $p\sim 15$. Furthermore, the thresholded mixer only leads to a small decrease in the probability to be found in the false minimum. This suggests that the effects of fluctuations is more complicated when realistic energy landscape structure is included, but that modifying the mixer to reduce the effect of fluctuations is still beneficial. Since the final diagonal elements of the Fubini-Study metric are increased by using the suppressed and thresholded strategy, as shown in figure  \ref{fig:gad_1_p_1_4_QGT_final}, a likely explanation is that the system ends the evolution in more complex states, but which still exhibit strong single qubit fluctuations on a subset of the qubits. 

The results in figure \ref{fig:gad_1_p_1_4} are demonstrated on a single instance of a QUBO constructed using the method described in \ref{Alg:problem_generation}. While these results are promising, it is worth examining some other instances to ensure that we have not accidentally selected an example with atypical behaviour. To understand typical behaviour, we consider multiple instances of problems generated in this way and compare the distributions. A convenient comparison method for our purposes is to compare the cumulative distribution function (CDF) of success probabilities. This allows for a comparison of the the methods ordered by how hard they are to solve using our methods, and therefore allows us to tell not only the typical behaviour, but also to detect, for example, if there is a small collection of atypically hard or easy instances which behave differently.

\begin{figure*}[htp!]
\begin{subfigure}[b]{0.35\textwidth}
	\begin{centering}
	\includegraphics[height=0.65\textwidth]{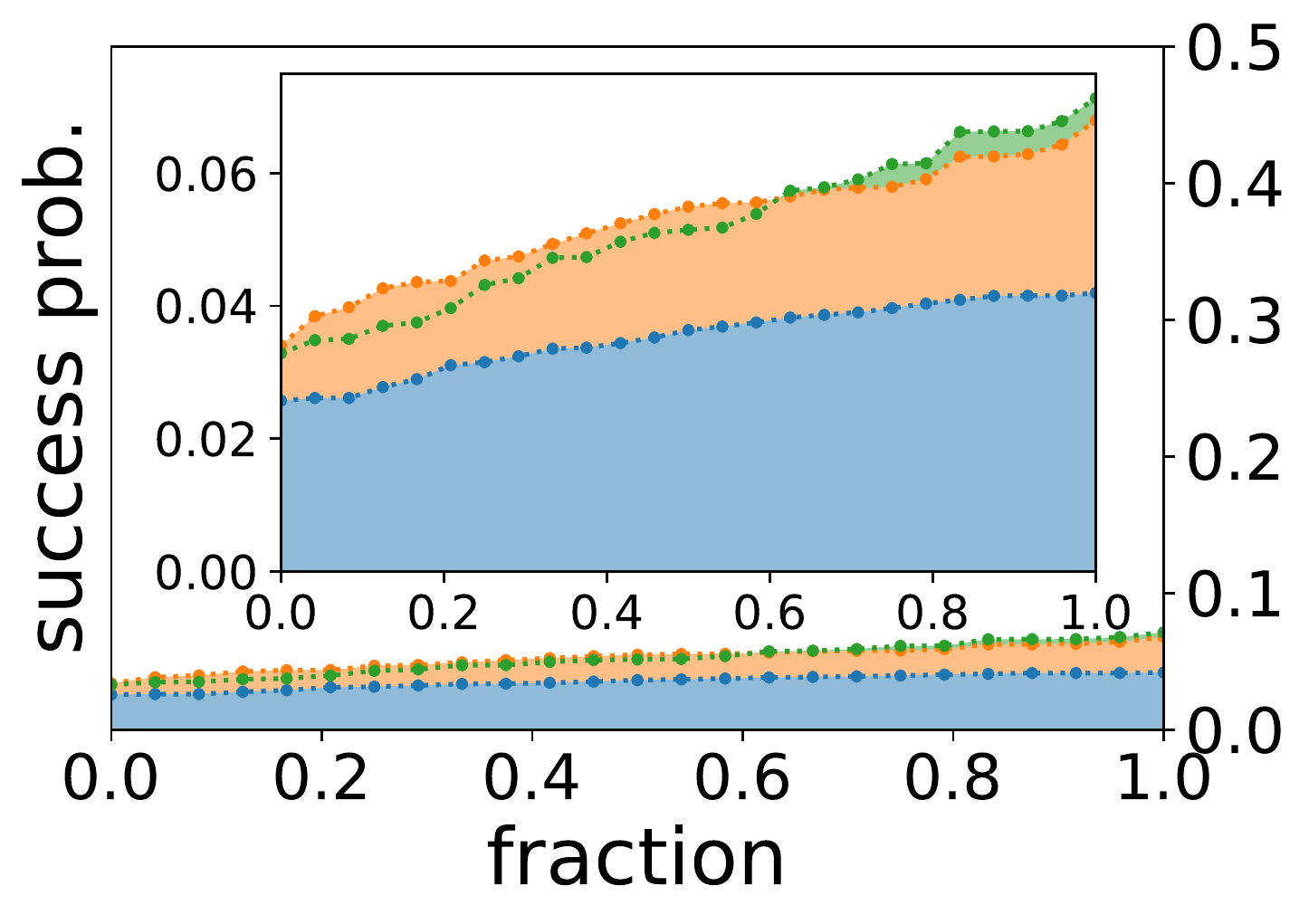}
	\par
	\end{centering}
	\caption{p = 10}
	\label{fig:433_varying_p10}
\end{subfigure}
\hspace{-0.04 \textwidth}
\begin{subfigure}[b]{0.35\textwidth}
	\begin{centering}
	\includegraphics[height=0.65\textwidth]{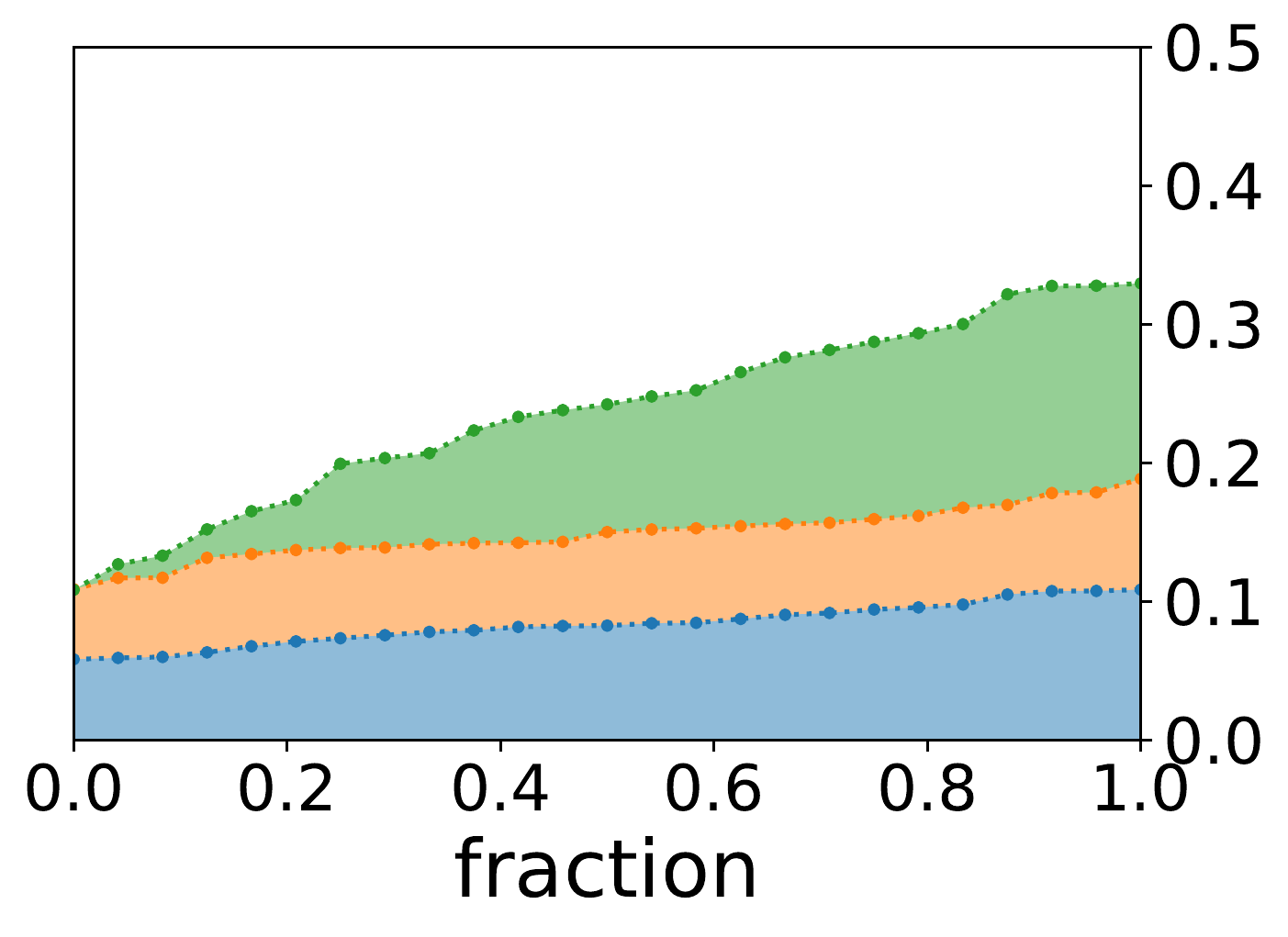}
	\par
	\end{centering}
	\caption{p = 30}
	\label{fig:433_varying_p30}
\end{subfigure}
\hspace{-0.045 \textwidth}
\begin{subfigure}[b]{0.35 \textwidth}
\centering
	\begin{centering}
	\includegraphics[height=0.65\textwidth]{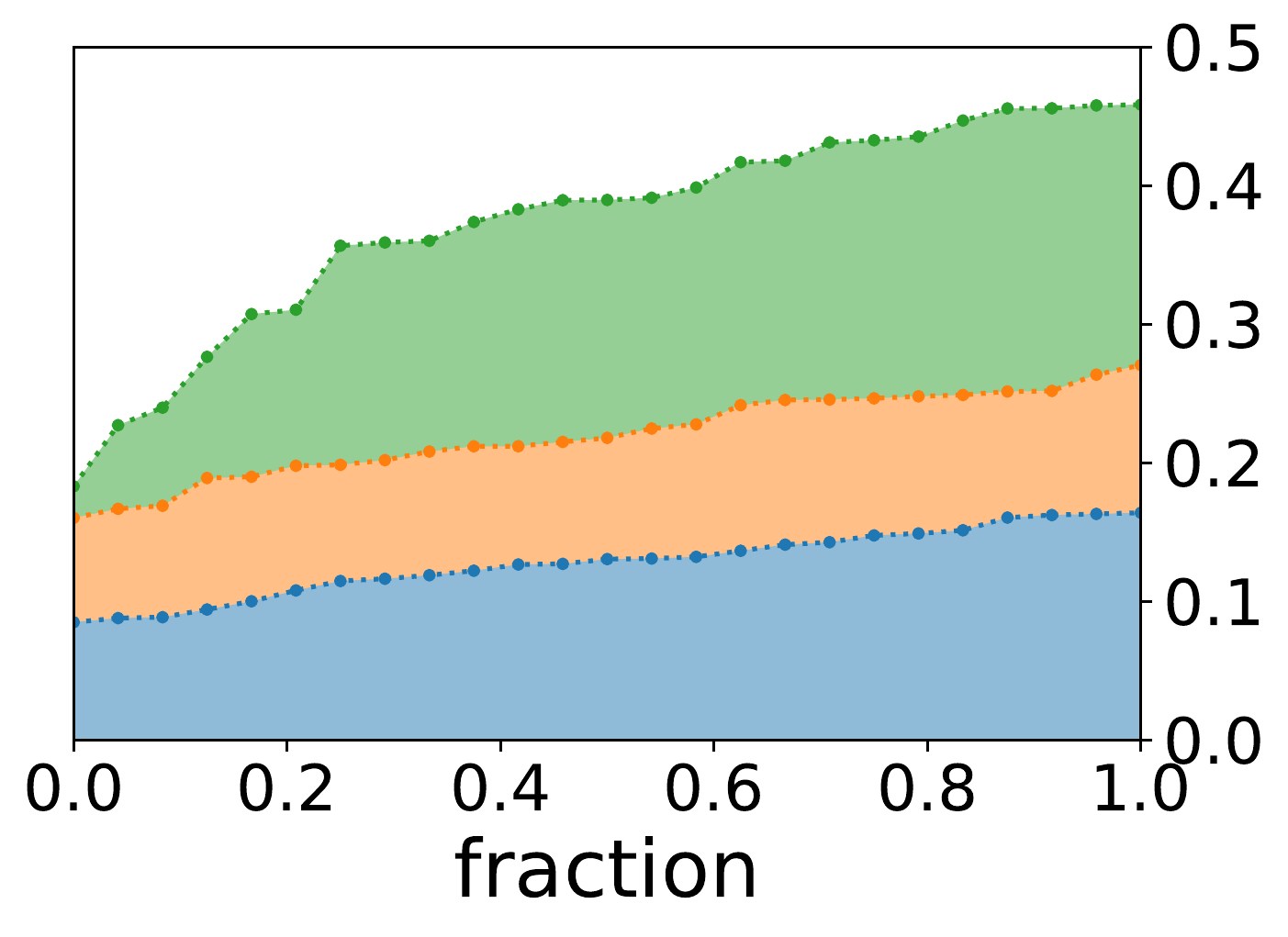}
	\par
	\end{centering}
	\caption{p  = 50}
	\label{fig:433_varying_p50}
\end{subfigure}
\caption{Cumulative distribution function (fraction of runs, with $25$ total) of success probability at the end of AQA solving a 10 variable problem generated using algorithm \ref{Alg:problem_generation} with $n_\mathrm{cut}=4$, $n_\mathrm{gadget}=3$, $J_{gadget}=0.25$, $J_{\mathrm{couple}}=0.5$, $\mathrm{Bias}=1.5$ for (a) p = 10, (b) p = 30, and (c) p = 50.  Blue corresponds to an unmodified mixer, while gold corresponds to the suppressed mixer and green to a thresholded mixer. Shading is included as a guide; all figures are plotted with the same y-axis scale for comparison. An inset with an expanded y-axis has been added to the first figure to show details.\label{fig:433_varying} }
\end{figure*}

As figure \ref{fig:433_varying} illustrates, especially at higher $p$ values, modification of the mixer leads to much better performance. Moreover, we can see that while figure \ref{fig:433_varying_p10} at $p=10$, the thresholded and suppressed strategies seem to be comparable, as $p$ is increased in figures \ref{fig:433_varying_p30} and \ref{fig:433_varying_p50}, there is a clear separation for all difficulties of problems in favor of the thresholded method. In fact for $p=50$, the lowest success probability instance performs better than the instance with the best performance under the unmodified protocol.

\begin{figure*}[htp!]
\begin{subfigure}[b]{0.35\textwidth}
	\begin{centering}
	\includegraphics[width=0.95\textwidth]{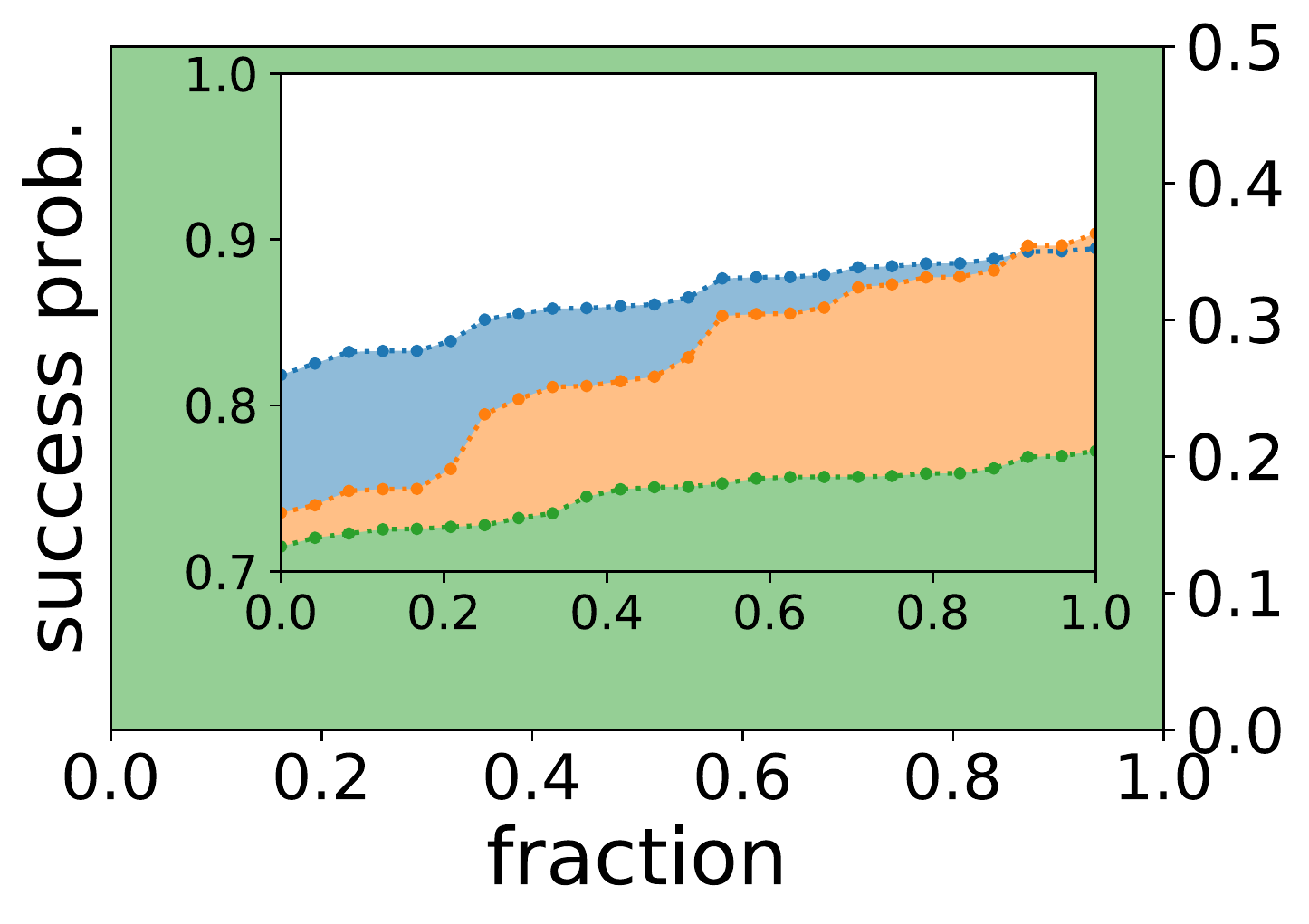}
	\par
	\end{centering}
	\caption{$n_\mathrm{cut}=2$, $n_\mathrm{gadget}=1$}
	\label{fig:size211_p50_runs25}
\end{subfigure}
\hspace{-0.04 \textwidth}
\begin{subfigure}[b]{0.35\textwidth}
	\begin{centering}
	\includegraphics[width=0.95\textwidth]{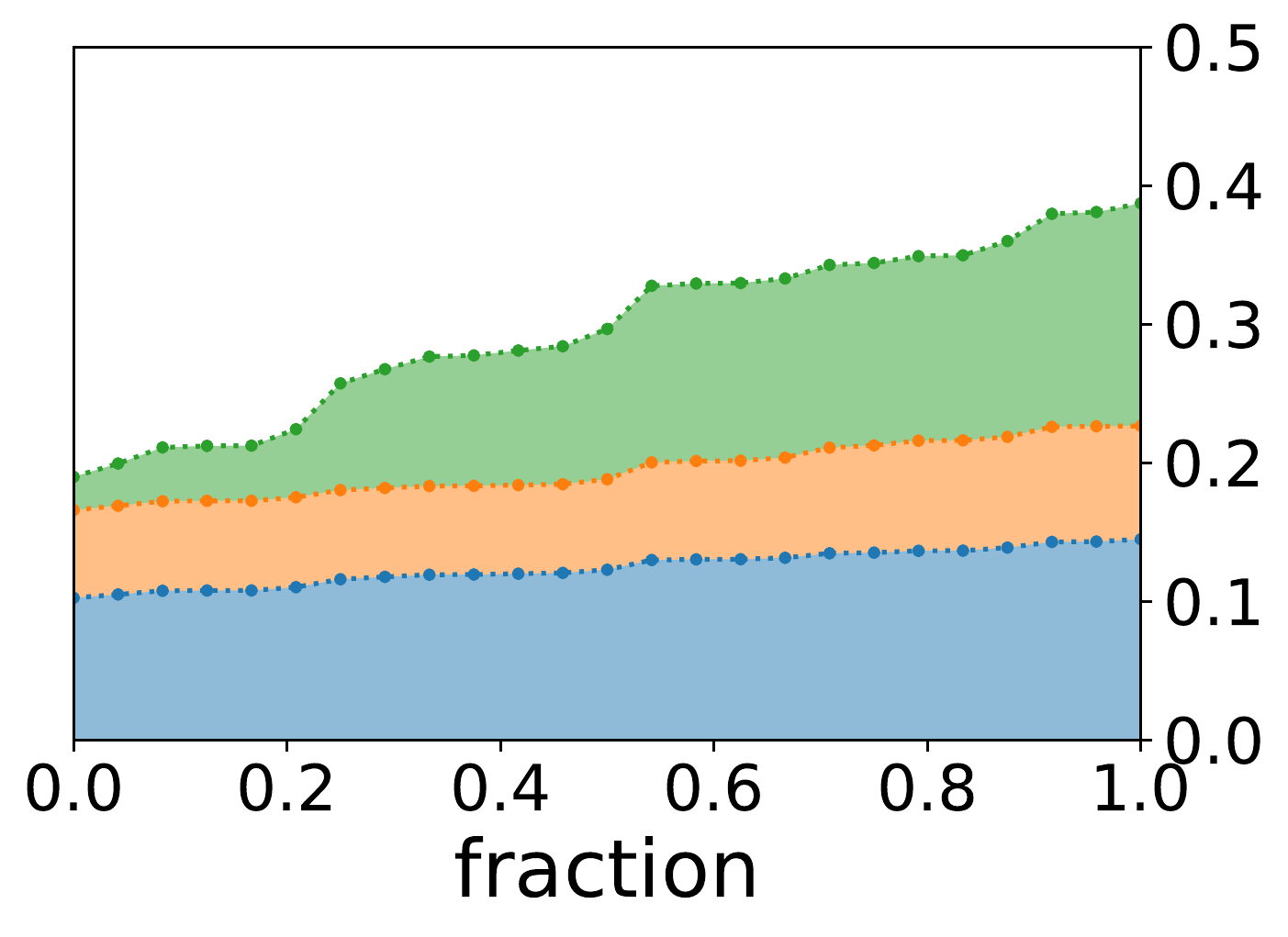}
	\par
	\end{centering}
	\caption{$n_\mathrm{cut}=2$, $n_\mathrm{gadget}=3$}
	\label{fig:size333_p50_runs25}
\end{subfigure}
\hspace{-0.04 \textwidth}
\begin{subfigure}[b]{0.35\textwidth}
	\begin{centering}
	\includegraphics[width=0.95\textwidth]{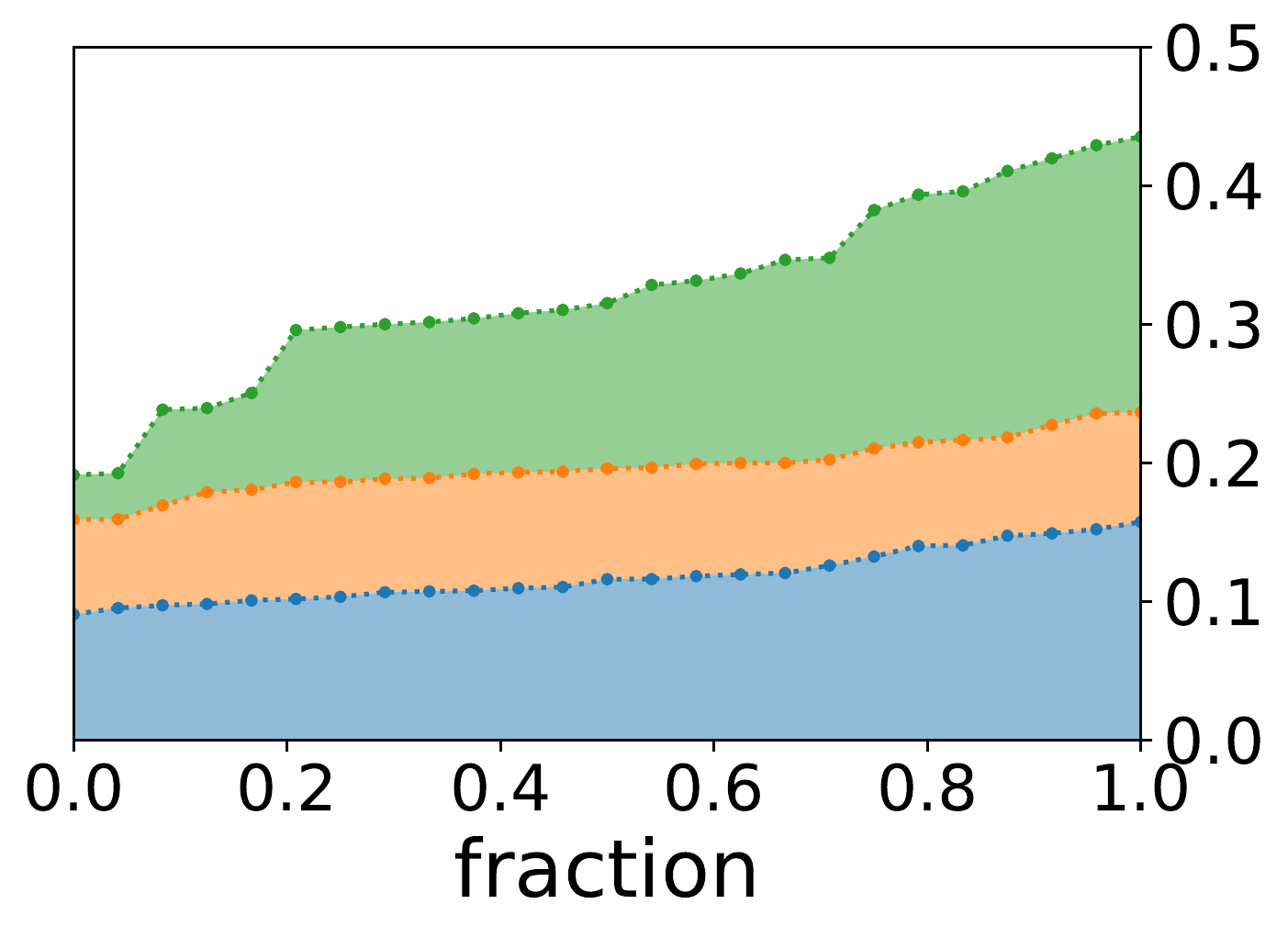}
	\par
	\end{centering}
	\caption{$n_\mathrm{cut}=3$, $n_\mathrm{gadget}=3$}
	\label{fig:size333_p50_runs25}
\end{subfigure}
\caption{Cumulative distribution function (fraction of runs, with 25 total) of success probability at the end of AQA solving a 10 variable problem generated using algorithm \ref{Alg:problem_generation} with  $J_{gadget}=0.25$, $J_{\mathrm{couple}}=0.5$, $\mathrm{Bias}=1.5$ and (a) $n_\mathrm{cut}=2$, $n_\mathrm{gadget}=1$, (b) $n_\mathrm{cut}=2$, $n_\mathrm{gadget}=3$, and (c) $n_\mathrm{cut}=3$, $n_\mathrm{gadget}=3$,  Blue corresponds to an unmodified mixer, while gold corresponds to the suppressed mixer and green to a thresholded mixer. Note that figure \ref{fig:433_varying_p50} is effectively a continuation to larger sizes. Shading is included as a guide; all figures are plotted with the same y-axis scale for comparison. An inset with an expanded y-axis has been added to the first figure to show details. \label{fig:CDF_size}}
\end{figure*}

Furthermore, we can see from figure \ref{fig:CDF_size} (and \ref{fig:433_varying_p50}) that the separation in performance seems to scale as a function of size with the effect becoming more pronounced as the problem QUBOs become bigger. 

\subsection{Random QUBO\label{sub:rand_qubo}}

\begin{figure*}[htp!]
\begin{subfigure}[b]{0.475\textwidth}
	\begin{centering}
 	\includegraphics[width=\textwidth]{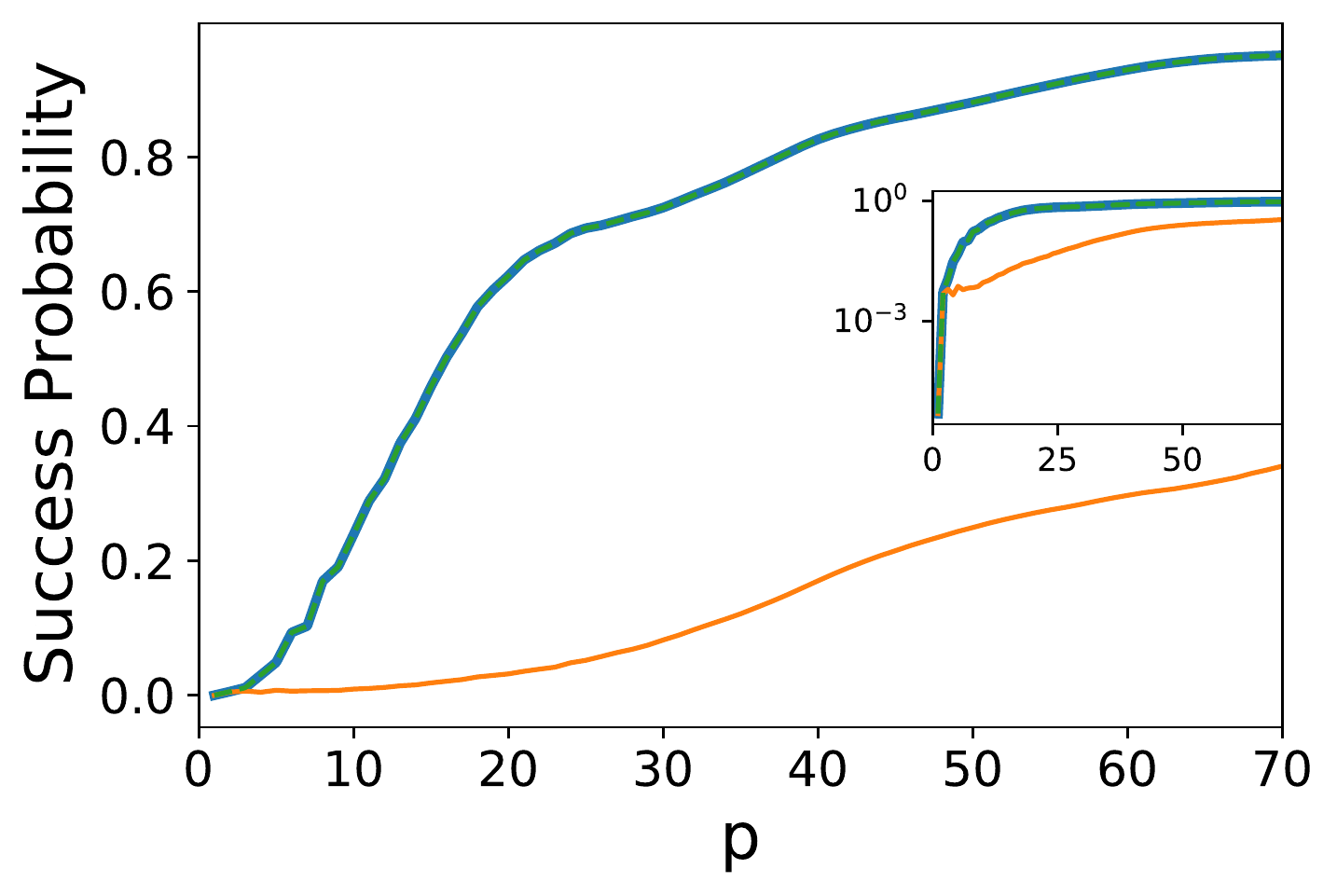}
	\par
	\end{centering}
	\caption{Success probabilities}
	\label{fig:rand_16_p_1_4_success}
\end{subfigure}
\begin{subfigure}[b]{0.475\textwidth}
	\begin{centering}
	\includegraphics[width=\textwidth]{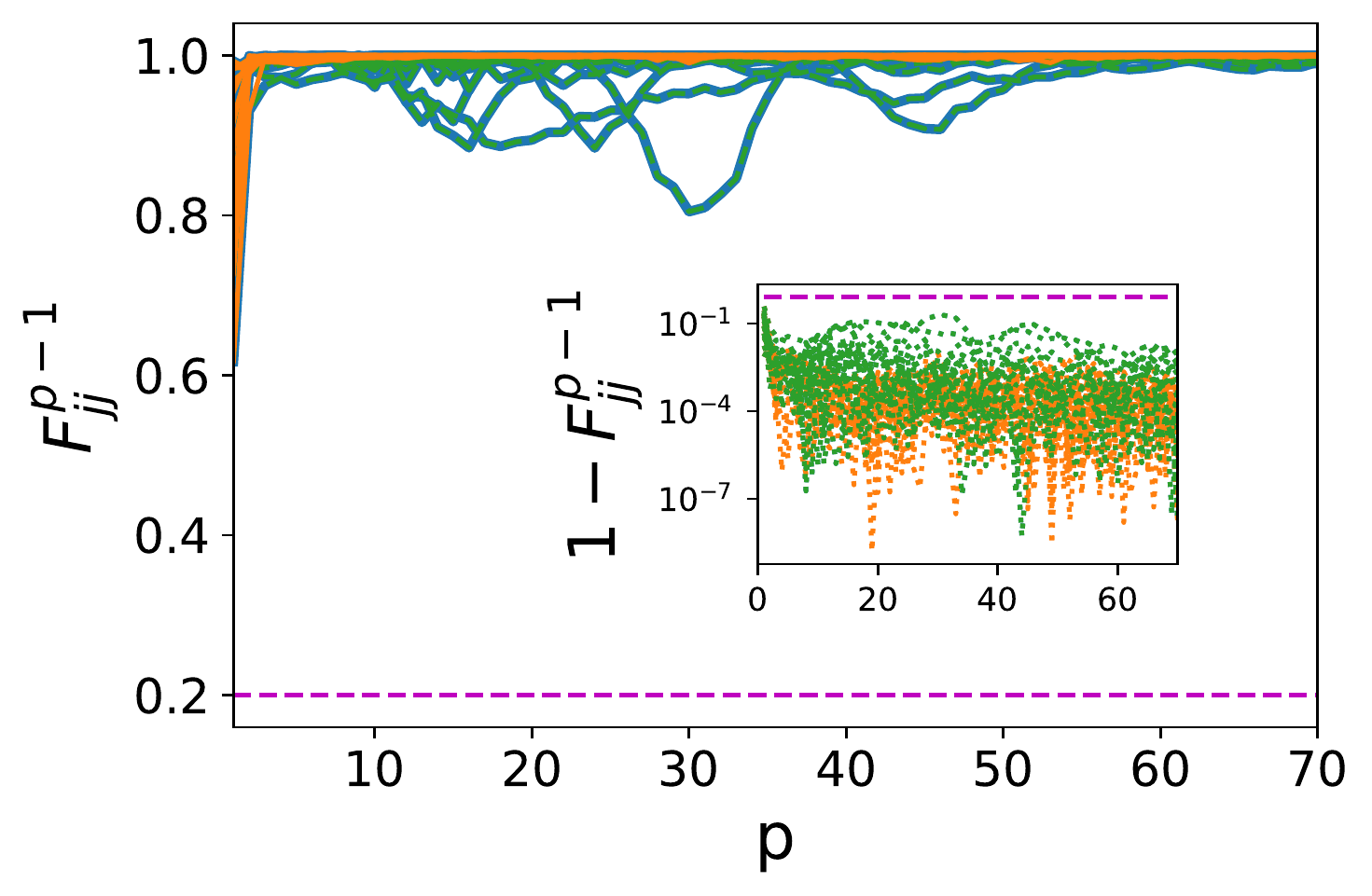}
	\par
	\end{centering}
	\caption{Metric elements}
	\label{fig:rand_16_p_1_4_QGT_final}
\end{subfigure}

\caption{Success probability (a) and diagonal elements of the Fubini Study-metric (b) for Pauli X operators at the end of QAOA at the end of AQA solving a 16 variable QUBO with matrix elements chosen uniformly at random. Each curve corresponds to a different metric element. Blue corresponds to an unmodified mixer, while gold corresponds to the suppressed mixer and green to a thresholded mixer. Because the unmodified and thresholded curves completely overlap we have visually indicated this by making the thresholded curve dashed and increasing the width of the unmodified curve. Likewise we have made all curves dotted in the inset of (b) for increased visibility. The insets are the same as the main plots but on a logarithmic scale. The specific QUBO used to create these plots can be found in table \ref{tab:rand_QUBO} of appendix II. \label{fig:rand_16_p_1_4}}

\end{figure*}

 We now want to understand why the thresholded mixer out-performs the suppressed version for a randomised QUBO. To test this, we examine what happens when the three protocols are used on a problem which has the structure of a hard optimisation problem, but not one where there is an engineered false minimum with strong fluctuations. To do this, we generate a random QUBO of size $16$ with all elements chosen uniformly between $-1$ and $1$. We find, as depicted in figure \ref{fig:rand_16_p_1_4_success}, that on this QUBO the suppressed mixer performs substantially more poorly than the unmodified mixer. Since the final Fubini-Study matrix elements are all far above the threshold at all times for the unmodified mixer, the thresholded and unmodified approaches are identical as depicted in figure \ref{fig:rand_16_p_1_4_QGT_final}.

\begin{figure*}[htp!]
\begin{subfigure}[b]{0.475\textwidth}
	\begin{centering}
	\includegraphics[width=\textwidth]{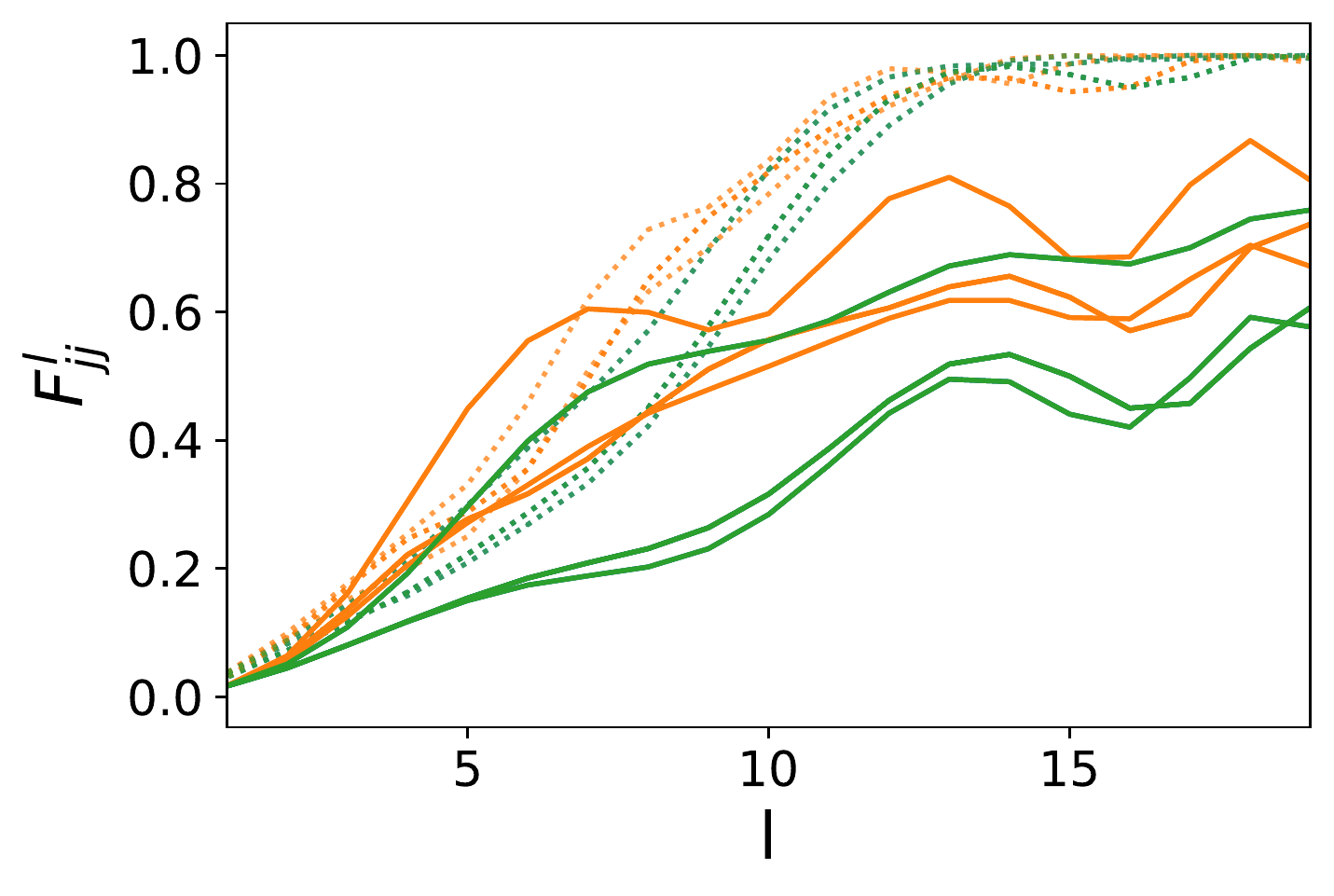}
	\par
	\end{centering}
	\caption{QUBO from \cite{dickson13a}}
	\label{fig:TA_1_p_1_4_slice_20}
\end{subfigure}
\begin{subfigure}[b]{0.475\textwidth}
	\begin{centering}
	\includegraphics[width=\textwidth]{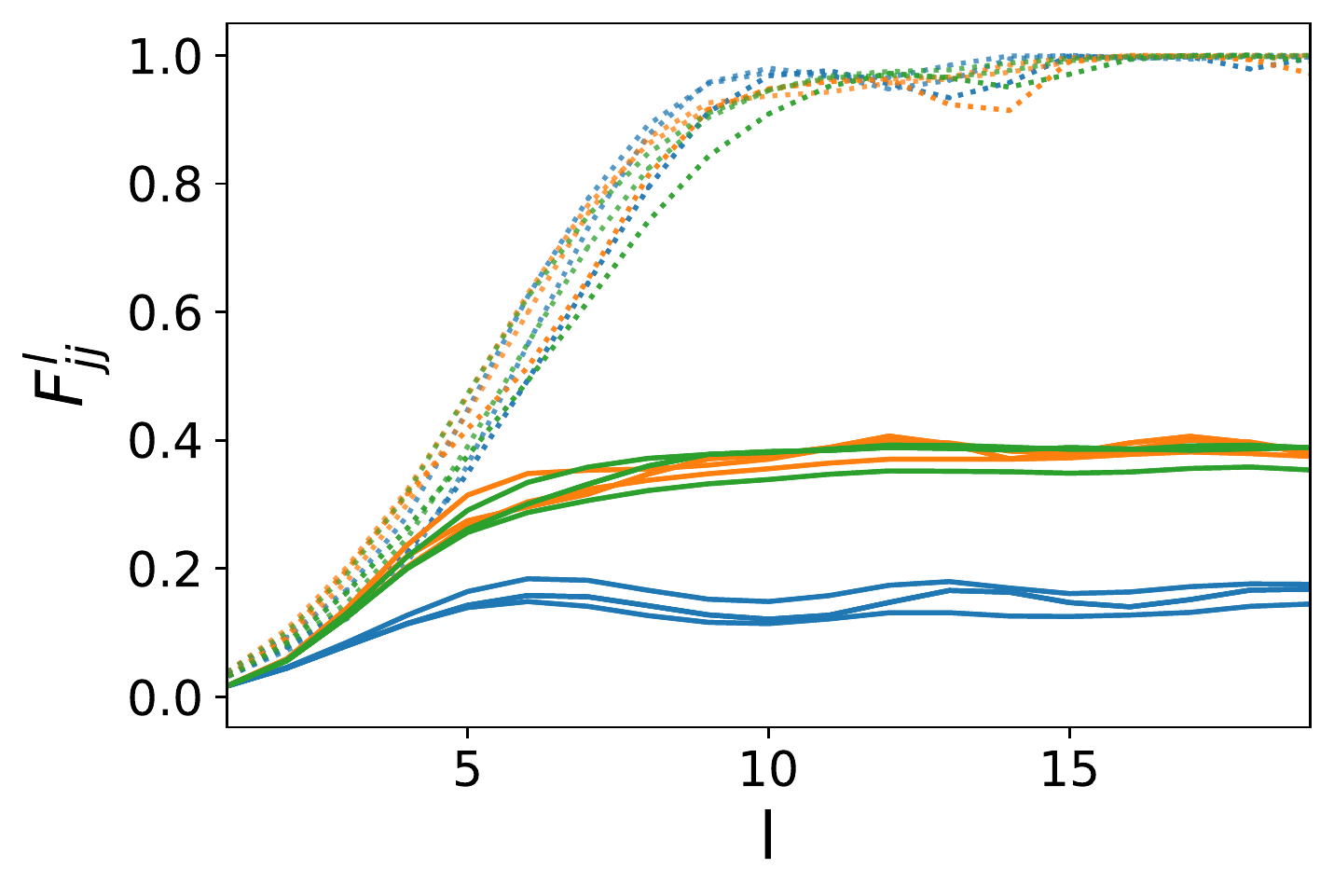}
	\par
	\end{centering}
	\caption{Modified QUBO from \cite{dickson13a}}
	\label{fig:TA_0p2_p_1_4_slice_20}
\end{subfigure}
\begin{subfigure}[b]{0.475\textwidth}
	\begin{centering}
	\includegraphics[width=\textwidth]{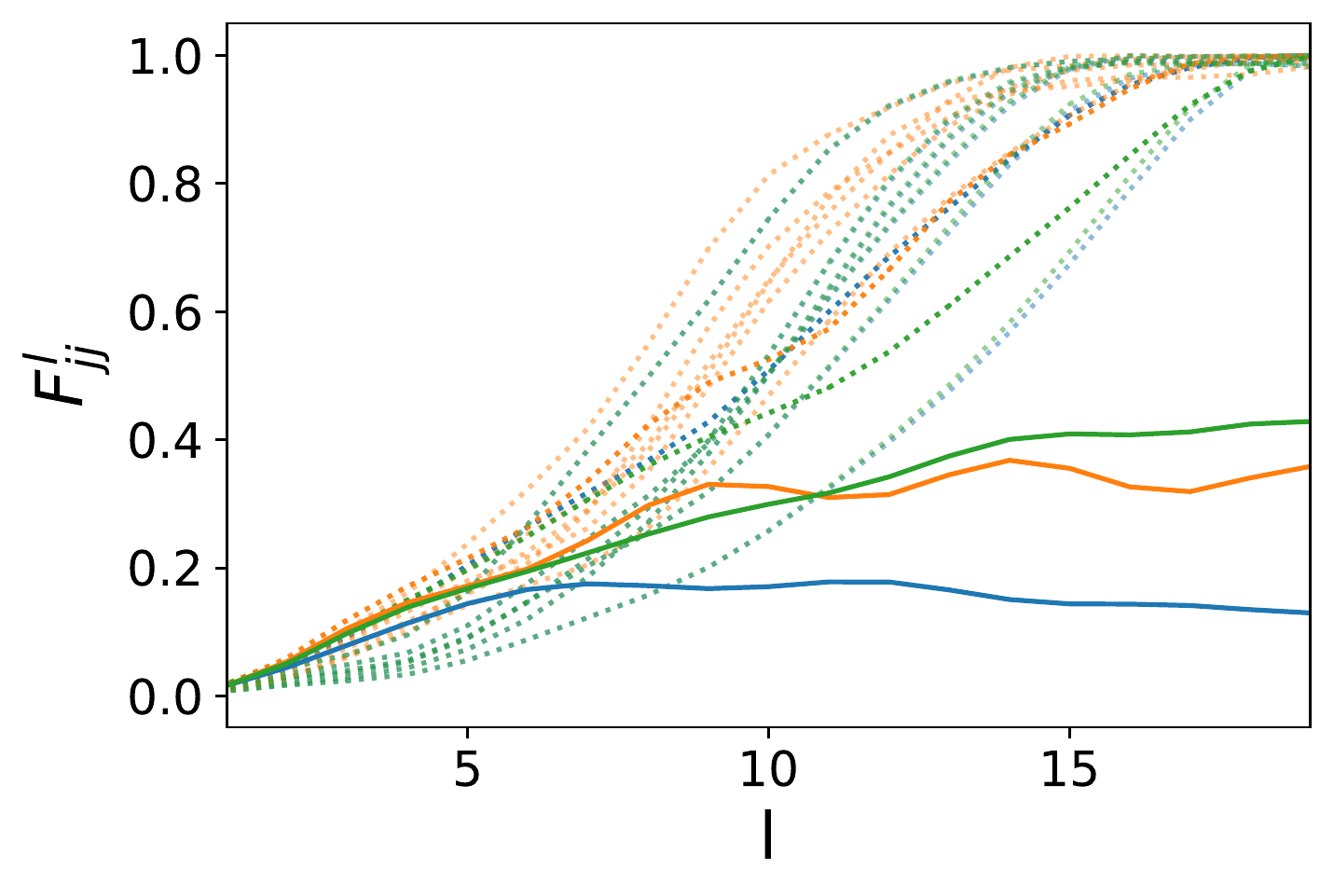}
	\par
	\end{centering}
	\caption{QUBO created using algorithm \ref{Alg:problem_generation}}
	\label{fig:gad_1_p_1_4_slice_20}
\end{subfigure}
\begin{subfigure}[b]{0.475\textwidth}
	\begin{centering}
	\includegraphics[width=\textwidth]{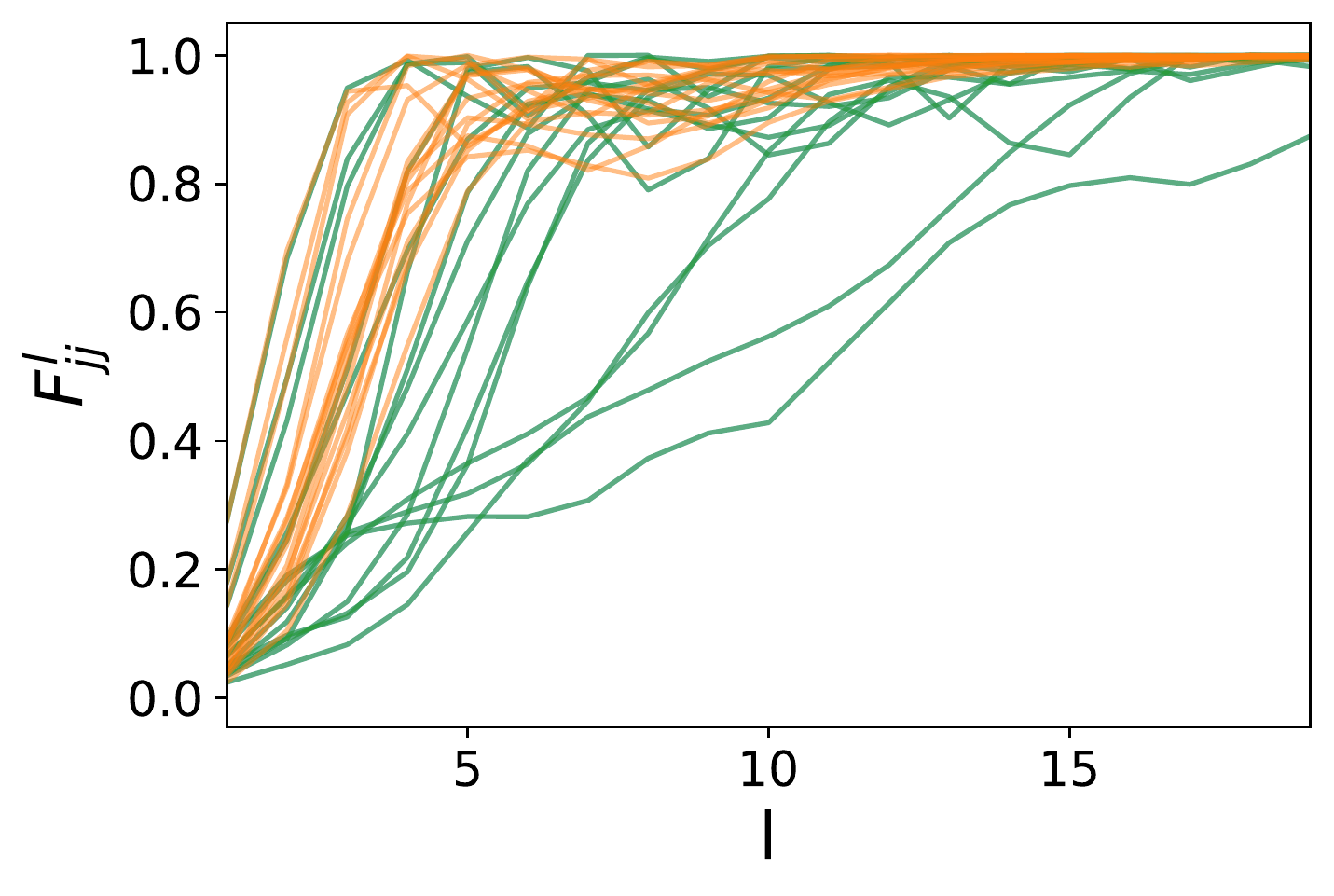}
	\par
	\end{centering}
	\caption{16 qubit random QUBO}
	\label{fig:rand_16_p_1_4_slice_20}
\end{subfigure}
\caption{Values of diagonal Pauli X Fubini-Study elements at different layers of the AQA protocol (indexed by $l$) for $p=20$ and various problem QUBOs. For all subfigures, blue corresponds to an unmodified mixer, while gold corresponds to the suppressed mixer and green to a thresholded mixer and $\tau(p)=\frac{\pi}{2p^{0.25}}$. In (a-c), elements corresponding to qubits which do not become effectively free within the false minimum are shown as fainter dot-dashed lines. The QUBO in (a) is the same as was used in figures \ref{fig:TA_1_p_1_4_success} and \ref{fig:TA_1_p_1_4_success} within section \ref{sub:Dickson_orig}, while the one in (b) was used in \ref{fig:TA_0p2_p_1_4_success}, \ref{fig:TA_0p2_p_1_4_false}, and \ref{fig:TA_0p2_p_1_4_QGT_final} within section \ref{sub:Dickson_mod}, (c) in \ref{fig:gad_1_p_1_4_success}, \ref{fig:gad_1_p_1_4_false}, and \ref{fig:gad_1_p_1_4_QGT_final} within section \ref{sub:spec_qubo}, the specific QUBO used here can be found in table \ref{tab:spec_QUBO}, and (d) in \ref{fig:rand_16_p_1_4_success} and \ref{fig:rand_16_p_1_4_QGT_final} within section \ref{sub:rand_qubo}, the specific QUBO used here can be found in table \ref{tab:rand_QUBO}. \label{fig:slice_20_summary}}
\end{figure*}

To understand why the suppressed mixer performs worse, it is useful to examine how the diagonal elements of the Fubini-Study metric behave throughout the QAOA protocol. This, in turn, will show the behavior of the terms of the mixer when the suppressed strategy is used. The various subfigures of figure \ref{fig:slice_20_summary} provide an indication as to why. For all QUBOs, the diagonal elements of the Fubini-Sudy metric show some non-monotonic behaviour, which, in the continuous time limit, leads to violation of the conditions for energy reduction stated in \cite{Callison21a}. This behaviour is likely to be sub-optimal and therefore detrimental to performance. We further note that, particularly in figures \ref{fig:TA_1_p_1_4_slice_20} and \ref{fig:gad_1_p_1_4_slice_20}, there seems to be stronger oscillations for the suppressed strategy, suggesting some kind of feedback effect from modifying elements of the mixer. Fortunately, as we have seen previously, the thresholded method seems to mostly avoid these detrimental effects by only changing mixer elements where necessary.

\section{Methods}

\subsection{Numerical \label{sec:num_meth}}

The numerical analysis presented in this paper was mostly performed using Python\cite{van2003python} while Matlab was used for some calculations. The cirq package was used extensively for gate-model simulations \cite{cirq}. QAOA was performed using scipy minimize COBYLA optimizer with tolerance set to zero to ensure that the optimizer ran to complete convergence or to the maximum number of iterations which was set to 300 \cite{2020SciPy-NMeth}. The base cirq simulator was used to sample 1000 solutions for each iteration of QAOA. For some deeper circuits a precision increase by setting the datatype to ``complex128'' was needed for the calculations to complete without errors, otherwise default settings were used. We also made extensive use of the numpy module\cite{numpy} for numerical calculations, matplotlib\cite{hunter2007matplotlib} for plotting and data visualisation, as well as jupyter notebooks \cite{jupyter,perez2007ipython}.

\subsection{Experimental \label{sec:exp_meth}}

Experiments were performed on an IonQ Harmony device in October 2022. For each point we performed $10$ separate experimental runs with $100$ samples each, error bars are the standard error between these runs. We found this yielded more stable results than a single run with $1000$ samples, possibly due to systematic errors in each run. Measurements of qubit $2$ in the $\{\ket{+},\ket{-}\}$ basis were accomplished by performing a Hardamard and then measuring in the Z basis.

\section{Discussion and Conclusions}

In this work, we have demonstrated a way in which the diagonal elements of the Fubini-Study metric can be used to moderate the size of mixer angles within QAOA and AQA protocols to suppress unwanted fluctuations which can lead the system to become trapped in false minima. This effect is well known under the context of quantum annealing but is not well explored in a gate-model setting. We find that while scaling the mixer angles according to the diagonal elements does work to suppress fluctuations, it is also detrimental to the overall performance of the algorithm. We find that this can be rectified by first performing a traditionally formulated QAOA or AQA run to identify which variables to target based on a threshold. This method, based on a threshold, both preserves the performance of the original algorithm on problems where fluctuations do not drive the system toward a false minima and suppresses unwanted fluctuations in cases where they do. We find that the advantages from our methods not only show up in the limit of large $p$, but that effects can be seen at moderate values between $10$ and $20$. This suggests that these methods will not only be relevant in the long term, when universal gate model machines can provide nearly flawless simulations of quantum annealing, but in the nearer term, where circuits must remain relatively shallow. While this work only examines relatively simple methods based on the diagonal elements of the Fubini-Study metric, the datasets from the measurements could be used to calculate the off diagonal elements as well, and more sophisticated control protocols could yield a further advantage. In particular, better protocols could be possible if fluctuations in the metric elements, and therefore the driver strength, could be suppressed. 

Our methods are complementary to other uses of the Fubini-Study metric, such as preconditioning, and act on degrees of freedom which are available but typically not used within QAOA (and which are not generally available for annealing). For this reason, these methods will be compatible with a large number of techniques including, but not limited to, quantum natural gradients \cite{Stokes2020quantumnatural} and warm starts for QAOA based on AQA schedules \cite{Willsch2022AQA}. While we have developed simple techniques which work well, it is possible that even more advanced techniques, for example using machine learning or taking advantage of the information stored in the off-diagonal elements of the Fubini-Study metric, may provide further enhancements to performance.

\section{Acknowledgements}

All authors were entirely supported by Quantum Computing Inc.~in completing this work aside from IonQ machine time. The authors thank IonQ for providing access to their machines to perform the experiments reported here. The authors thank Uchenna Chukwu and Daiwei Zhou for useful discussions.

\bibliography{reference}  

\section*{Appendix I: QAOA convergence}

\begin{figure}[!ht]
\begin{centering}
\includegraphics[width=7cm]{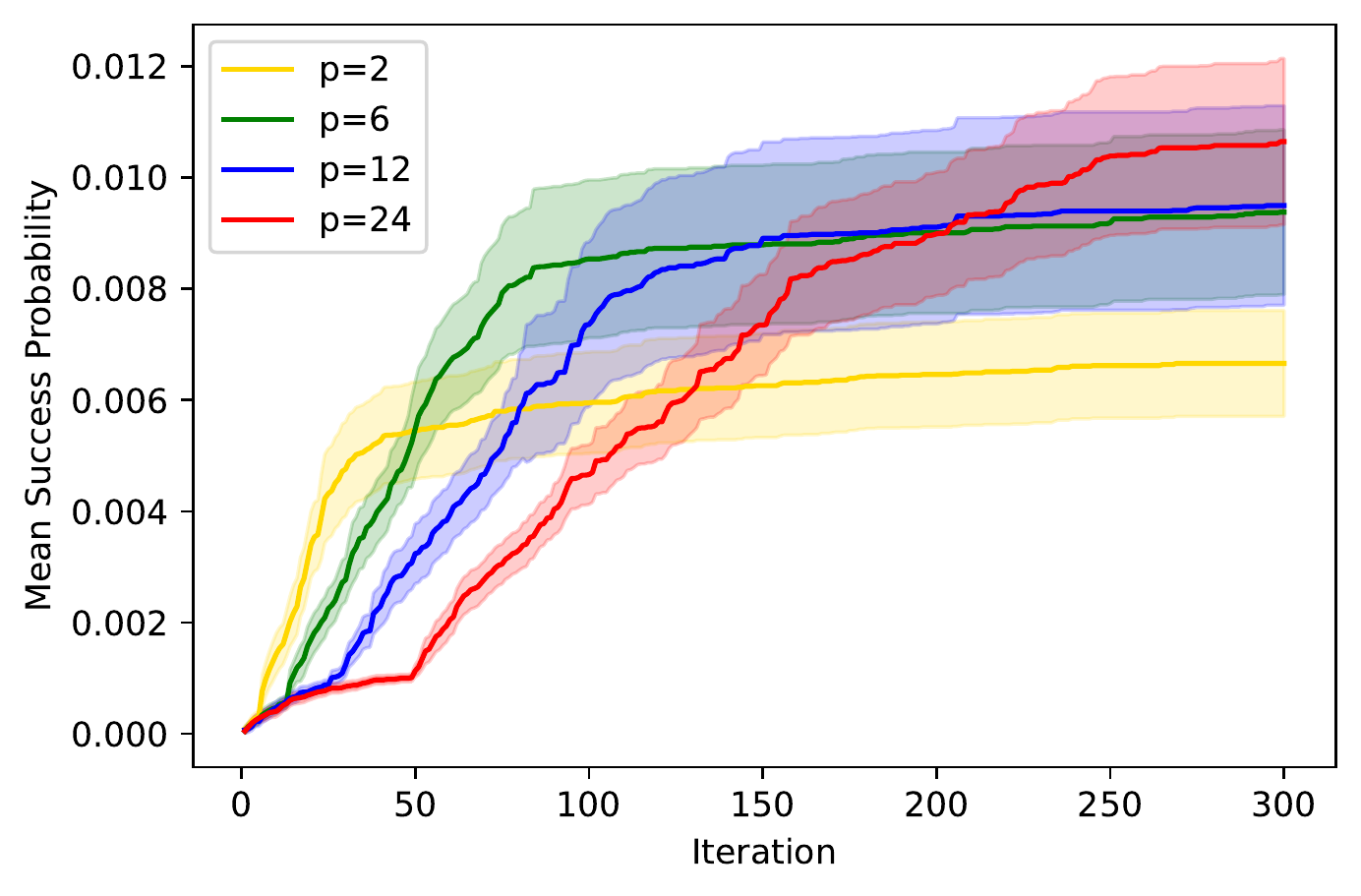}
\par
\end{centering}

\caption{Convergence of mean success probability QAOA (using methods described in section \ref{sec:num_meth}) for different $p$ values and the same problem as used in figure \ref{fig:gad_1_p_1_4_success} (and appearing in table \ref{tab:spec_QUBO}) . Shaded area indicates two standard errors above and below the mean of 100 runs for each of the 300 iterations 
\label{fig:QAOA_converge}}
\end{figure}

\begin{figure}[!ht]
\begin{centering}
\includegraphics[width=7cm]{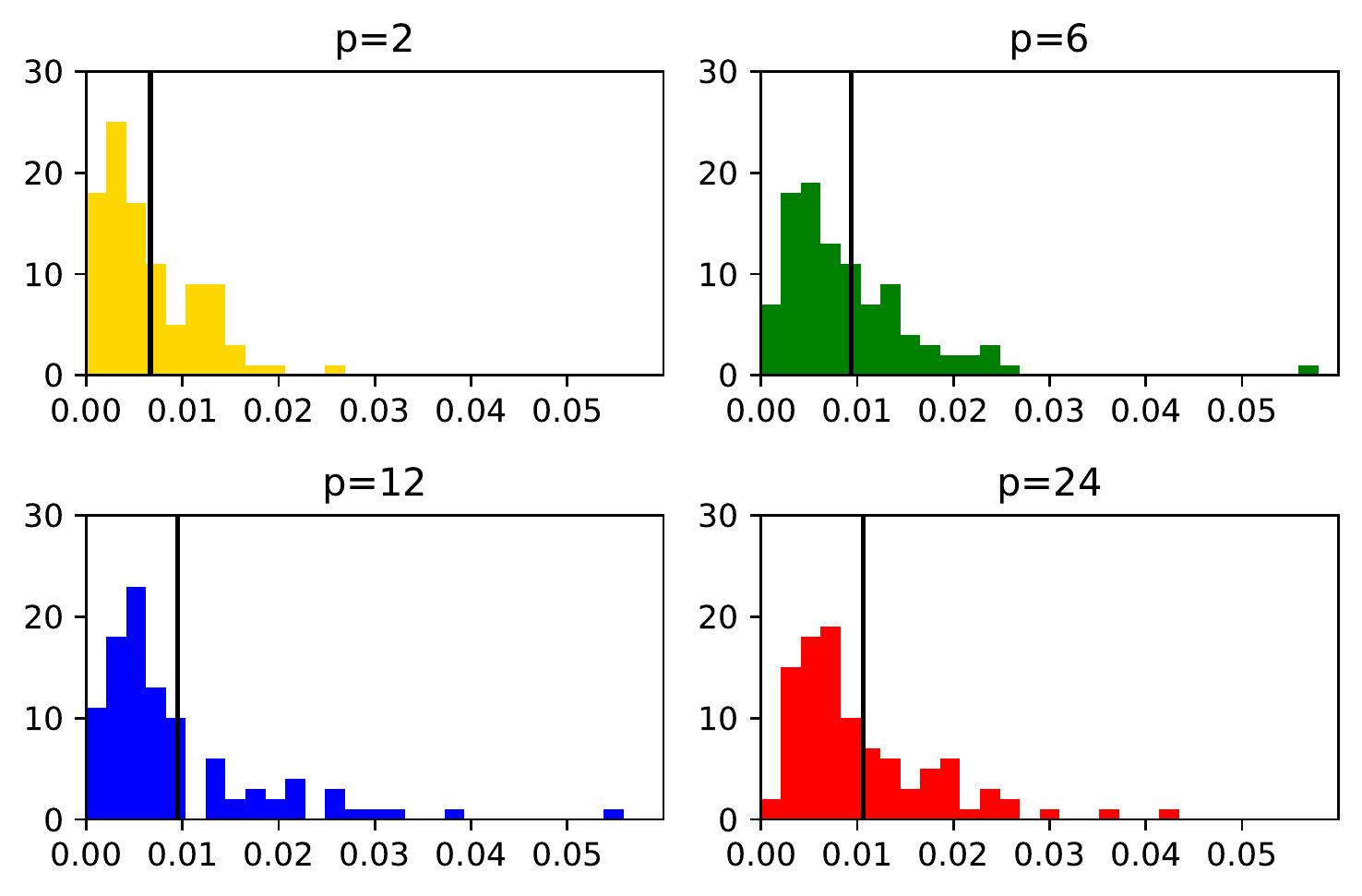}
\par
\end{centering}
\caption{Histogram of counts by success probability for final iteration of QAOA for each depth $p$}
\label{fig:qaoa_last_iter_hist}
\end{figure}

When comparing to an optimised method such as QAOA, it is important to verify that the method is converging correctly. We have therefore included figure \ref{fig:QAOA_converge}, which shows the mean success probability, when running QAOA on the same problem used in figure \ref{fig:gad_1_p_1_4_success}, is indeed converging. The shaded area around each line marking two standard errors above and below the mean shows the likely range of values for the mean success probability of each group. To create the line for the mean success probability, each simulation contributed its highest success probability that it had obtained on any iteration up to and including the current iteration's result. To illustrate the distribution of the results, the highest success probability for each run is shown below (figure \ref{fig:qaoa_last_iter_hist})

\section*{Applendix II: Randomly generated QUBOs used in examples}

For reproducibility purposes, the QUBO to generate figures \ref{fig:gad_1_p_1_4}, \ref{fig:gad_1_p_1_4_QGT_final}, \ref{fig:QAOA_converge}, and \ref{fig:gad_1_p_1_4_slice_20}, generated using algorithm \ref{Alg:problem_generation} can be found in table \ref{tab:spec_QUBO}. While the random QUBO used in figures \ref{fig:rand_16_p_1_4} and \ref{fig:rand_16_p_1_4_slice_20} appears in table \ref{tab:rand_QUBO}.

\begin{sidewaystable}
{\scriptsize
\begin{tabular}{cccccccccccccc} 
 0.654888 & -0.155566 & 0.580933 & 0.691021 & -0.516198 & -0.921744 & -0.020833 & -0.020833 & -0.020833 & -0.020833 & 0.0 & 0.0 & 0.0 & 0.0\tabularnewline 
-0.155566 & 0.85708 & -0.114969 & -0.201265 & -0.734799 & 0.682853 & -0.020833 & -0.020833 & -0.020833 & -0.020833 & 0.0 & 0.0 & 0.0 & 0.0\tabularnewline 
0.580933 & -0.114969 & -2.208083 & 0.434978 & 0.689647 & 0.284161 & 0.020833 & 0.020833 & 0.020833 & 0.020833 & 0.0 & 0.0 & 0.0 & 0.0\tabularnewline 
0.691021 & -0.201265 & 0.434978 & -1.700257 & 0.880849 & -0.43866 & 0.020833 & 0.020833 & 0.020833 & 0.020833 & 0.0 & 0.0 & 0.0 & 0.0\tabularnewline 
-0.516198 & -0.734799 & 0.689647 & 0.880849 & -0.847565 & 0.861399 & -0.020833 & -0.020833 & -0.020833 & -0.020833 & 0.0 & 0.0 & 0.0 & 0.0\tabularnewline 
-0.921744 & 0.682853 & 0.284161 & -0.43866 & 0.861399 & -0.801343 & 0.020833 & 0.020833 & 0.020833 & 0.020833 & 0.0 & 0.0 & 0.0 & 0.0\tabularnewline 
-0.020833 & -0.020833 & 0.020833 & 0.020833 & -0.020833 & 0.020833 & 0.75 & -0.25 & -0.25 & -0.25 & -1.0 & 0.0 & 0.0 & 0.0\tabularnewline 
-0.020833 & -0.020833 & 0.020833 & 0.020833 & -0.020833 & 0.020833 & -0.25 & 0.75 & -0.25 & -0.25 & 0.0 & -1.0 & 0.0 & 0.0\tabularnewline 
-0.020833 & -0.020833 & 0.020833 & 0.020833 & -0.020833 & 0.020833 & -0.25 & -0.25 & 0.75 & -0.25 & 0.0 & 0.0 & -1.0 & 0.0\tabularnewline 
-0.020833 & -0.020833 & 0.020833 & 0.020833 & -0.020833 & 0.020833 & -0.25 & -0.25 & -0.25 & 0.75 & 0.0 & 0.0 & 0.0 & -1.0\tabularnewline 
0.0 & 0.0 & 0.0 & 0.0 & 0.0 & 0.0 & -1.0 & 0.0 & 0.0 & 0.0 & 2.0 & 0.0 & 0.0 & 0.0\tabularnewline 
0.0 & 0.0 & 0.0 & 0.0 & 0.0 & 0.0 & 0.0 & -1.0 & 0.0 & 0.0 & 0.0 & 2.0 & 0.0 & 0.0\tabularnewline 
0.0 & 0.0 & 0.0 & 0.0 & 0.0 & 0.0 & 0.0 & 0.0 & -1.0 & 0.0 & 0.0 & 0.0 & 2.0 & 0.0\tabularnewline 
0.0 & 0.0 & 0.0 & 0.0 & 0.0 & 0.0 & 0.0 & 0.0 & 0.0 & -1.0 & 0.0 & 0.0 & 0.0 & 2.0\tabularnewline 
\end{tabular} 
}
    \caption{The QUBO to generate figures \ref{fig:gad_1_p_1_4}, \ref{fig:gad_1_p_1_4_QGT_final}, \ref{fig:QAOA_converge}, and \ref{fig:gad_1_p_1_4_slice_20}, generated using algorithm \ref{Alg:problem_generation}}
    \label{tab:spec_QUBO}
\end{sidewaystable}

\begin{sidewaystable}
{\tiny
\begin{tabular}{cccccccccccccccc} 
 -0.50716 & -0.626067 & -0.370979 & 0.204478 & -0.731826 & 0.497795 & -0.089543 & 0.778661 & -0.51009 & -0.799264 & 0.433472 & -0.299784 & 0.603779 & -0.604381 & -0.570997 & -0.724014\tabularnewline 
-0.626067 & -0.881469 & 0.413613 & 0.710529 & -0.007755 & -0.851639 & 0.50867 & 0.490394 & -0.826299 & 0.385482 & 0.887413 & -0.415068 & -0.161579 & -0.588002 & -0.137618 & 0.897788\tabularnewline 
-0.370979 & 0.413613 & 0.173839 & 0.163411 & 0.858976 & 0.955427 & 0.030875 & -0.492069 & 0.051477 & 0.774003 & 0.551384 & -0.387463 & -0.889746 & -0.075674 & 0.526891 & 0.240785\tabularnewline 
0.204478 & 0.710529 & 0.163411 & 0.540605 & 0.212848 & -0.621795 & 0.730394 & 0.139587 & -0.978558 & -0.284671 & 0.750477 & 0.812303 & 0.287821 & -0.405717 & -0.501293 & 0.753629\tabularnewline 
-0.731826 & -0.007755 & 0.858976 & 0.212848 & 0.742537 & 0.544606 & -0.265008 & 0.728864 & -0.152381 & -0.470731 & -0.503601 & -0.391257 & 0.203381 & 0.356628 & -0.833427 & -0.797992\tabularnewline 
0.497795 & -0.851639 & 0.955427 & -0.621795 & 0.544606 & -0.287941 & -0.730018 & 0.077989 & 0.390509 & -0.942507 & -0.715465 & 0.15538 & 0.198981 & -0.149508 & -0.175403 & 0.943773\tabularnewline 
-0.089543 & 0.50867 & 0.030875 & 0.730394 & -0.265008 & -0.730018 & 0.257362 & 0.585086 & 0.953278 & -0.931662 & 0.456079 & -0.954151 & 0.742424 & -0.74155 & 0.82253 & -0.395456\tabularnewline 
0.778661 & 0.490394 & -0.492069 & 0.139587 & 0.728864 & 0.077989 & 0.585086 & -0.430128 & -0.835616 & 0.709582 & 0.516953 & -0.732259 & 0.996284 & -0.977999 & 0.210182 & -0.111627\tabularnewline 
-0.51009 & -0.826299 & 0.051477 & -0.978558 & -0.152381 & 0.390509 & 0.953278 & -0.835616 & -0.359618 & -0.935286 & 0.969805 & -0.063232 & 0.783269 & 0.104272 & -0.304731 & -0.236273\tabularnewline 
-0.799264 & 0.385482 & 0.774003 & -0.284671 & -0.470731 & -0.942507 & -0.931662 & 0.709582 & -0.935286 & -0.450217 & 0.237596 & -0.391438 & -0.828199 & 0.240247 & 0.223207 & -0.344709\tabularnewline 
0.433472 & 0.887413 & 0.551384 & 0.750477 & -0.503601 & -0.715465 & 0.456079 & 0.516953 & 0.969805 & 0.237596 & 0.154905 & 0.423843 & -0.110097 & 0.65209 & 0.385553 & 0.135372\tabularnewline 
-0.299784 & -0.415068 & -0.387463 & 0.812303 & -0.391257 & 0.15538 & -0.954151 & -0.732259 & -0.063232 & -0.391438 & 0.423843 & 0.694109 & -0.01228 & -0.729243 & 0.565898 & 0.747734\tabularnewline 
0.603779 & -0.161579 & -0.889746 & 0.287821 & 0.203381 & 0.198981 & 0.742424 & 0.996284 & 0.783269 & -0.828199 & -0.110097 & -0.01228 & -0.462504 & 0.209157 & -0.436286 & 0.697161\tabularnewline 
-0.604381 & -0.588002 & -0.075674 & -0.405717 & 0.356628 & -0.149508 & -0.74155 & -0.977999 & 0.104272 & 0.240247 & 0.65209 & -0.729243 & 0.209157 & 0.91753 & 0.116758 & 0.14103\tabularnewline 
-0.570997 & -0.137618 & 0.526891 & -0.501293 & -0.833427 & -0.175403 & 0.82253 & 0.210182 & -0.304731 & 0.223207 & 0.385553 & 0.565898 & -0.436286 & 0.116758 & -0.222499 & 0.515007\tabularnewline 
-0.724014 & 0.897788 & 0.240785 & 0.753629 & -0.797992 & 0.943773 & -0.395456 & -0.111627 & -0.236273 & -0.344709 & 0.135372 & 0.747734 & 0.697161 & 0.14103 & 0.515007 & -0.228068\tabularnewline 
\end{tabular} 
}
    \caption{The randomly generated QUBO to generate figures \ref{fig:rand_16_p_1_4} and \ref{fig:rand_16_p_1_4_slice_20}.} 
    \label{tab:rand_QUBO}
\end{sidewaystable}

\end{document}